\documentclass[12pt,prd, aps, showpacs, superscriptaddress,floatfix]{revtex4}
\usepackage{graphicx}
\usepackage{amsmath,bm}
\usepackage{amssymb}
\usepackage{caption}
\usepackage{subcaption}
\usepackage{epsfig, rotate, epsf,setspace, amssymb, amsxtra, placeins,slashed,simplewick}
\newcommand{\MeV}{\mathrm{MeV}}
\newcommand{\s}[1]{\slash \!\!\! #1}
\newcommand{\MSbar}{$\overline{\mathrm{MS}}$~}

\def\simle{
    \mathrel{\rlap{\raise 0.511ex 
        \hbox{$<$}}{\lower 0.511ex \hbox{$\sim$}}}}
\DeclareMathOperator*{\SumInt}{%
\mathchoice%
  {\ooalign{$\displaystyle\sum$\cr\hidewidth$\displaystyle\int$\hidewidth\cr}}
  {\ooalign{\raisebox{.14\height}{\scalebox{.7}{$\textstyle\sum$}}\cr\hidewidth$\textstyle\int$\hidewidth\cr}}
  {\ooalign{\raisebox{.2\height}{\scalebox{.6}{$\scriptstyle\sum$}}\cr$\scriptstyle\int$\cr}}
  {\ooalign{\raisebox{.2\height}{\scalebox{.6}{$\scriptstyle\sum$}}\cr$\scriptstyle\int$\cr}}
}
\newcommand\riken{RIKEN-BNL Research Center, Brookhaven National
  Laboratory, Upton, NY 11973, USA}
\newcommand\bnl{Brookhaven National Laboratory, Upton, NY 11973, USA}
\newcommand\edinb{SUPA, School of Physics, The University of
  Edinburgh, Edinburgh EH9 3JZ, UK}
\newcommand{\camb}{DAMTP, University of Cambridge, Wilberforce Road, Cambridge CB3 0WA, UK}
\newcommand\cu{Physics Department, Columbia University, New York,
  NY 10027, USA}
\newcommand\uconn{Physics Department, University of Connecticut,
  Storrs, CT 06269-3046, USA}
\newcommand\soton{School of Physics and Astronomy, University of
  Southampton,  Southampton SO17 1BJ, UK}

\newcommand{\glasgow}{SUPA, School of Physics and Astronomy, University of Glasgow, Glasgow, G12 8QQ, UK}
\newcommand{\plymouth}{School of Computing and Mathematics, Plymouth University
Plymouth, PL4 8AA, UK}

\begin{document}
\title{$\mathbf{K\rightarrow \pi\pi}$ $\mathbf{\Delta I = 3/2}$ decay amplitude in the continuum limit.}

\author{T.Blum}\affiliation{\uconn}\affiliation{\riken}
\author{P.A.Boyle}\affiliation{\edinb}
\author{N.H.Christ}\affiliation{\cu}
\author{J.Frison}\affiliation{\edinb}
\author{N.Garron}\affiliation{\camb}\affiliation{\plymouth}
\author{T.Janowski}\affiliation{\soton}
\author{C.Jung}\affiliation{\bnl}
\author{C.Kelly}\affiliation{\riken}
\author{C.Lehner}\affiliation{\bnl}
\author{A.Lytle}\affiliation{\glasgow}
\author{R.D.~Mawhinney}\affiliation{\cu}
\author{C.T.Sachrajda}\affiliation{\soton}
\author{A.Soni}\affiliation{\bnl}
\author{H.Yin}\affiliation{\cu}
\author{D.Zhang}\affiliation{\cu}
\collaboration{RBC and UKQCD Collaborations}

\begin{abstract}
We present new results for the amplitude $A_2$ for a kaon to decay into two pions with isospin $I=2$: Re$A_2 = 1.50(4)_\mathrm{stat}(14)_\mathrm{syst}\times 10^{-8}\ {\rm GeV}$;  Im$A_2 = -6.99(20)_\mathrm{stat}(84)_\mathrm{syst}\times 10^{-13}\ {\rm GeV}$.
These results were obtained from two ensembles generated at physical quark masses (in the isospin limit) with inverse lattice spacings $a^{-1}=1.728(4)$\,GeV and $2.358(7)$\,GeV. We are therefore able to perform a continuum extrapolation and hence largely to remove the dominant  systematic uncertainty from our earlier results~\cite{{Blum:2011ng,Blum:2012uk}}, that due to lattice artifacts. 
The only previous lattice computation of $K\to\pi\pi$ decays at physical kinematics was performed using an ensemble at a single, rather coarse, value of the lattice spacing [$a^{-1}\simeq 1.37(1)\,$GeV]. We confirm the observation reported in~\cite{Boyle:2012ys} that there is a significant cancellation between the two dominant contributions to Re$A_2$ which we suggest is an important ingredient in understanding the $\Delta I=1/2$ rule, Re$A_0$/Re$A_2\simeq 22.5$, where the subscript denotes the total isospin of the two-pion final state. Our result for $A_2$ implies that the electroweak penguin contribution to $\epsilon^\prime/\epsilon$ is Re($\epsilon^\prime/\epsilon)_\textrm{EWP}=-(6.6\pm 1.0)\times 10^{-4}$.
\end{abstract}
\maketitle
\section{Introduction}\label{sec:intro}

Nonleptonic $K\to\pi\pi$ decays continue to be an important class of processes in the phenomenology of the standard model of particle physics. Historically it was in these decays that both direct and indirect $CP$-violation were discovered and the challenges for theoretical physicists include an explanation of the long-standing puzzle of the $\Delta I=1/2$ rule and an \textit{ab initio} computation of  $\epsilon^\prime/\epsilon$. Developments in the theoretical framework of lattice QCD and in efficient algorithms, together with the availability of  the latest computing power, have made meeting these challenges feasible. A significant element of the current joint research program of the RBC and UKQCD collaborations is the evaluation of the $K\to\pi\pi$ amplitudes $A_0$ and $A_2$, where the subscript represents the isospin of the two-pion final state (which by Bose symmetry is restricted to 0 or 2). In this paper we present our latest results for $A_2$.

In \cite{Blum:2011ng,Blum:2012uk} we reported on the first results from a lattice determination of the amplitude $A_2$ for $K\to(\pi\pi)_{I=2}$ decays, where $I$ is the total isospin of the two-pion final state:
\begin{equation}\label{eq:a2old}
\mathrm{Re}A_2=1.381(46)_\mathrm{stat}(258)_\mathrm{syst}\,10^{-8}\,\mathrm{GeV},
\quad\mathrm{Im}A_2=-6.54(46)_\mathrm{stat}(120)_\mathrm{syst}\,10^{-13}\,\mathrm{GeV}\,.
\end{equation}
This was the first quantitative calculation of an amplitude for a realistic hadronic weak decay
and hence extended the framework of lattice simulations into the important domain of nonleptonic
weak decays.
As explained in the Introduction of~\cite{Blum:2012uk}, in order to obtain the result in Eq.\,(\ref{eq:a2old}) it was necessary to overcome a number of theoretical problems and exploit recent improvements in algorithms and the opportunities provided by increases in computing resources. The systematic errors in (\ref{eq:a2old}) are dominated by the fact that the calculation was performed at a single, rather coarse, value of the lattice spacing ($a\simeq 0.14\,$fm). We estimated these errors to be $O(15\%)$.

In this paper we repeat the calculation at two finer values of the lattice spacing and perform the continuum extrapolation.The simulations are carried out at physical pion masses (with unitary sea- and valence-quark masses) using our two new ensembles with lattice spacings  $a=0.011$\,fm and $a=0.084$\,fm. 
Our new result is presented in Eq.\,(\ref{eq:A2context}) and we reproduce it here for the reader's convenience:
\begin{equation}
\boxed{{\rm Re}(A_2) = 1.50(4)_\mathrm{stat}(14)_\mathrm{syst}\times 10^{-8}\ {\rm GeV};\quad
{\rm Im}(A_2) = -6.99(20)_\mathrm{stat}(84)_\mathrm{syst}\times 10^{-13}\ {\rm GeV}\,.}
\end{equation}

A very interesting feature of our earlier calculation of $A_2$ was the observation that the two dominant contributions to Re\,$A_2$ show a significant numerical cancellation~\cite{Boyle:2012ys}. We argued in \cite{Boyle:2012ys} that this cancellation is an important element in the explanation of the $\Delta I=1/2$ rule,  Re$A_0$/Re$A_2\simeq 22.5$. We confirm this cancellation in the present calculation. Of course, before we can claim that we fully understand the $\Delta I$=1/2 rule, we need to compute $A_0$ at physical quark masses and momenta; this calculation is even more challenging than the evaluation of $A_2$ but is under way. For the status of this calculation we refer the reader to~\cite{ckdaiqian}. 

The structure of the remainder of this paper is as follows. In the next section we present the parameters of the two ensembles used in this calculation. The evaluation of the bare matrix elements and the renormalization of the lattice operators are discussed in Secs.~\ref{sec:bare} and \ref{sec:NPR} respectively. We consider finite-volume effects in Sec.\,\ref{sec:FV} and present an overview of the different sources of systematic uncertainty in Sec.\,\ref{sec:errors}. We perform the continuum extrapolation in Sec.\,\ref{sec:contextrap} and present our final result in Eq.\,(\ref{eq:A2context}). Section \ref{sec:concs} contains our conclusions and a brief discussion of the prospects for the reduction of the errors in $A_2$ as well as for the calculation of $A_0$. There is one appendix in which we reproduce the calculation from~\cite{Lin:2002nq} of the Lellouch-L\"uscher factor for finite-volume corrections in the context of chiral perturbation theory. This calculation demonstrates how to disentangle
the finite-volume corrections which decrease exponentially with increasing lattice volume (a source
of systematic error) from those which decrease as a
power of the volume (which are corrected by
the Lellouch-L\"uscher factor). 
This calculation also clarifies a misunderstanding of these effects in the literature~
\cite{Aubin:2008vh}.
\section{Details of the simulation}\label{sec:simulations}
The calculations described below have been performed on two new 2+1~flavor ensembles generated with the Iwasaki gauge action and with M\"obius domain-wall fermions~\cite{Blum:2014tka}
. The parameters of the ensembles are
\begin{enumerate}
\item[(i)]$48^3\times 96 \times 24$ with $\beta = 2.13$ ($a^{-1}=1.728(4)$\,GeV);
\item[(ii)] $64^3 \times 128 \times 12$ with $\beta=2.25$ ($a^{-1}=2.357(7)$\,GeV).
\end{enumerate}
These two ensembles use the M\"obius variant of domain
wall fermions~\cite{Brower:2012vk} with a M\"obius
scale factor $\alpha=2$.
For compactness of notation we will refer to these ensembles as $48^3$ and $64^3$ respectively.
The lattice spacing and quark masses were set by choosing the masses of the pion, kaon and the $\Omega$-baryon to be equal to their physical values.
The corresponding sea-quark masses are $am_{ud} = 7.8 \times 10^{-4}$ and $am_s = 3.62\times 10^{-2}$, with the residual mass $am_{res} = 6.19(6)\times 10^{-4}$ for the $48^3$ ensemble and $am_{ud} = 6.78\times 10^{-4}$, $am_s = 2.661 \times 10^{-2}$ and $am_{res} = 2.93(8) \times 10^{-4}$ for the $64^3$ ensemble.
The two ensembles have approximately the same physical volume with spatial extent $L\simeq 5.5$\,fm, enabling the continuum extrapolation to be separated from finite-volume effects which we estimate separately. For more details on these ensembles see~\cite{Blum:2014tka} and we will return briefly to the determination of the lattice spacings in the context of the continuum extrapolation in Sec.\,\ref{sec:contextrap}.

The results presented below were obtained using 76 gauge configurations on the $48^3$ ensemble and 40 on the $64^3$ ensemble. 
The large statistical uncertainty one expects with a
relatively small number of gauge configurations can be
significantly reduced if we perform many measurements
on each configuration in which the sources and sinks
are simply translated in space and time~~\cite{Blum:2014tka}.  Performing
multiple measurements on the same configuration offers
two important opportunities for increased efficiency.
First if we can use a low-mode deflation method such as
eigCG~\cite{Stathopoulos:2007zi} we will be able to amortize the setup costs
of such an approach over a large number of inversions.  
Second we can use the all mode averaging technique \cite{Blum:2012uh} and
perform most of these many inversions at reduced precision
and use a relatively few accurate inversions to determine
a correction that guarantees systematic double precision
but with an additional (usually small) statistical error
that reflects the small number of accurate solves.  
Specifically for the $48^3$ ensemble, the eigCG method
was used in single precision with 600 approximate low-lying
eigenvectors and a stopping residual of $10^{-4}$. The approximate (wall source) propagators were computed on all 96 time slices. The accurate solves used to correct the approximation were computed on time slices 0, 76, 72, 68, 64, 60 and 56 with Conjugate Gradient (CG) stopping residual $10^{-8}$. (This choice of time-slice separations is not related to
the $K\to\pi\pi$ calculation presented here but to an
accompanying calculation of $B_K$~\cite{Blum:2014tka}.) To ensure that no bias results from the choice of
inexact solves for which the correction is calculated,
this complete pattern of source time slices for the
accurate solves was shifted by a different random time
displacement on each configuration.
A similar procedure was used on the $64^3$ ensemble but with 1500 low modes and a stopping residual of $10^{-5}$ for the approximate solves and accurate solves on time slices 0, 103, 98, 93, 88, 83, 78 and 73. On both ensembles, the accurate CG solves were also
computed using eigCG, exploiting the approximate
eigenvectors created during the inaccurate applications
of eigCG.

\begin{table}[t]
\begin{center}
\begin{tabular}{|c|c|c|c|c|}
	\hline
	& $m_\pi$ & $m_K$ & $E_{\pi\pi}$ & $m_K-E_{\pi\pi}$\\
	\hline
	$48^3$ (lattice units) & $8.050(13) \times 10^{-2}$ & $2.8867(15) \times 10^{-1}$ & $2.873(13) \times 10^{-1}$ & $1.4(14)\times 10^{-3}$ \\
	$64^3$ (lattice units) & $5.904(14) \times 10^{-2}$ & $2.1531(14) \times 10^{-1}$ & $2.1512(68) \times 10^{-1}$ & $9(10) \times 10^{-4}$\\
	\hline
	$48^3$ (MeV) & $139.1(2)$ & $498.82(26)$ & $496.5(16)$ & $2.4(24)$ \\
	$64^3$ (MeV) & $139.2(3)$ & $507.4(4)$ & $507.0(16)$ & $2.1(26)$\\
	\hline
	\end{tabular}
	\caption{Pion and kaon masses and the I=2 two-pion energies in lattice and physical units measured on the $48^3$ and $64^3$ ensembles. The momentum of each of the final-\rm{stat}e pions is $\pm\pi/L$ in each of the three spatial directions.}
	\label{tab:masse}\end{center}
\end{table}

Measurements on the $48^3$ and $64^3$ ensembles are separated by 20 and 40 molecular dynamics (MD) units respectively. In order to study the effects of autocorrelations we bin the data. We find that the effects are small, typically leading to a variation of the statistical errors of less than 10\%. The results presented below were obtained after binning the 76 configurations of the $48^3$ ensemble into 19 bins of 4 configurations and the 40 configurations of the $64^3$ ensemble into 8 bins of 5 configurations. The 40 configurations from the $64^3$ ensemble are precisely those used in the global analysis reported in~\cite{Blum:2014tka}. The 76 configurations from the $48^3$ ensemble include 73 of the 80 used in~\cite{Blum:2014tka}. We have however, repeated the relevant analysis of~\cite{Blum:2014tka}, including the determination of the lattice spacings, using precisely the 76 configurations for which we have computed $A_2$. This makes it possible to compute standard
jackknife errors for our physical results which
necessarily depend upon the value of the lattice
spacing.

\begin{figure}[t]
		\begin{center}
		\begin{tabular}{ccc}
		\includegraphics[width=0.445\hsize]{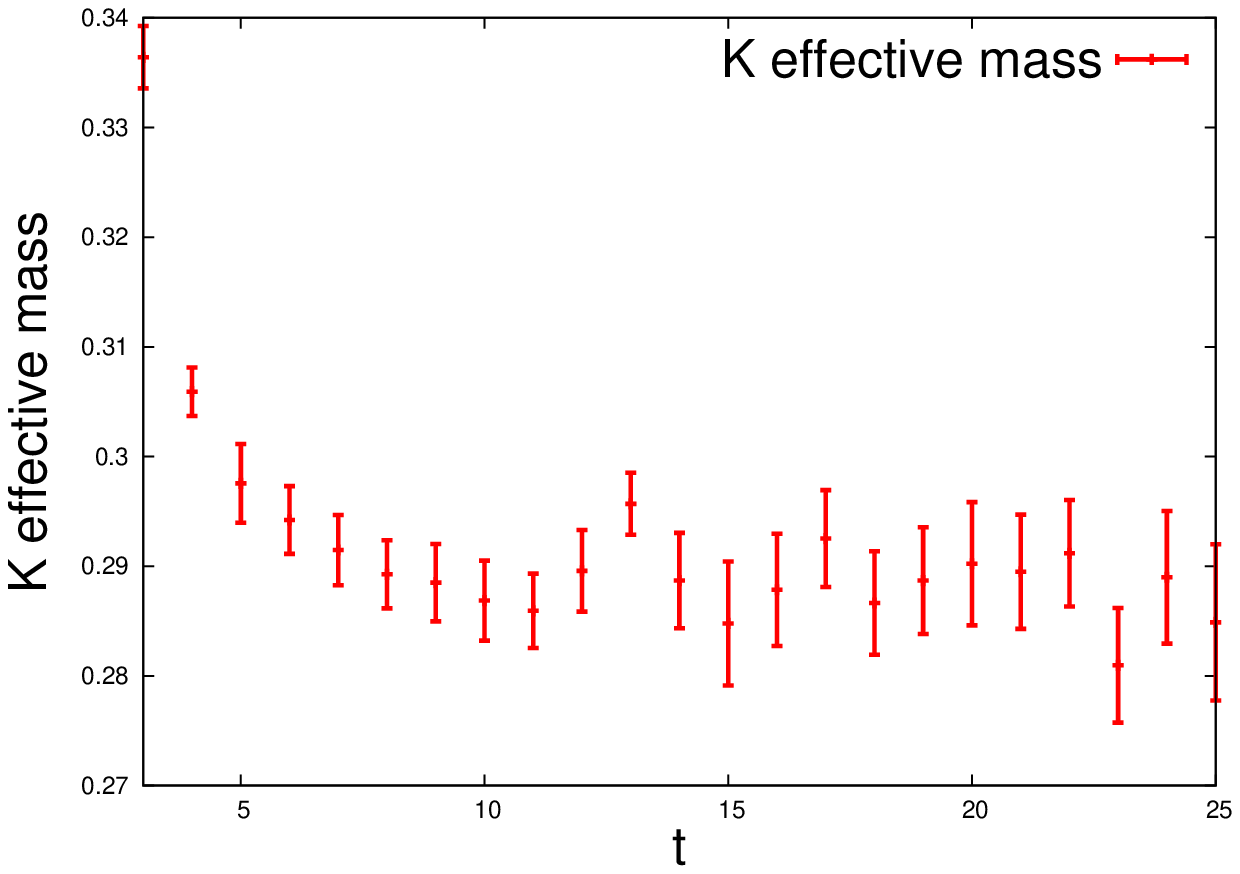} \qquad \includegraphics[width=0.445\hsize]{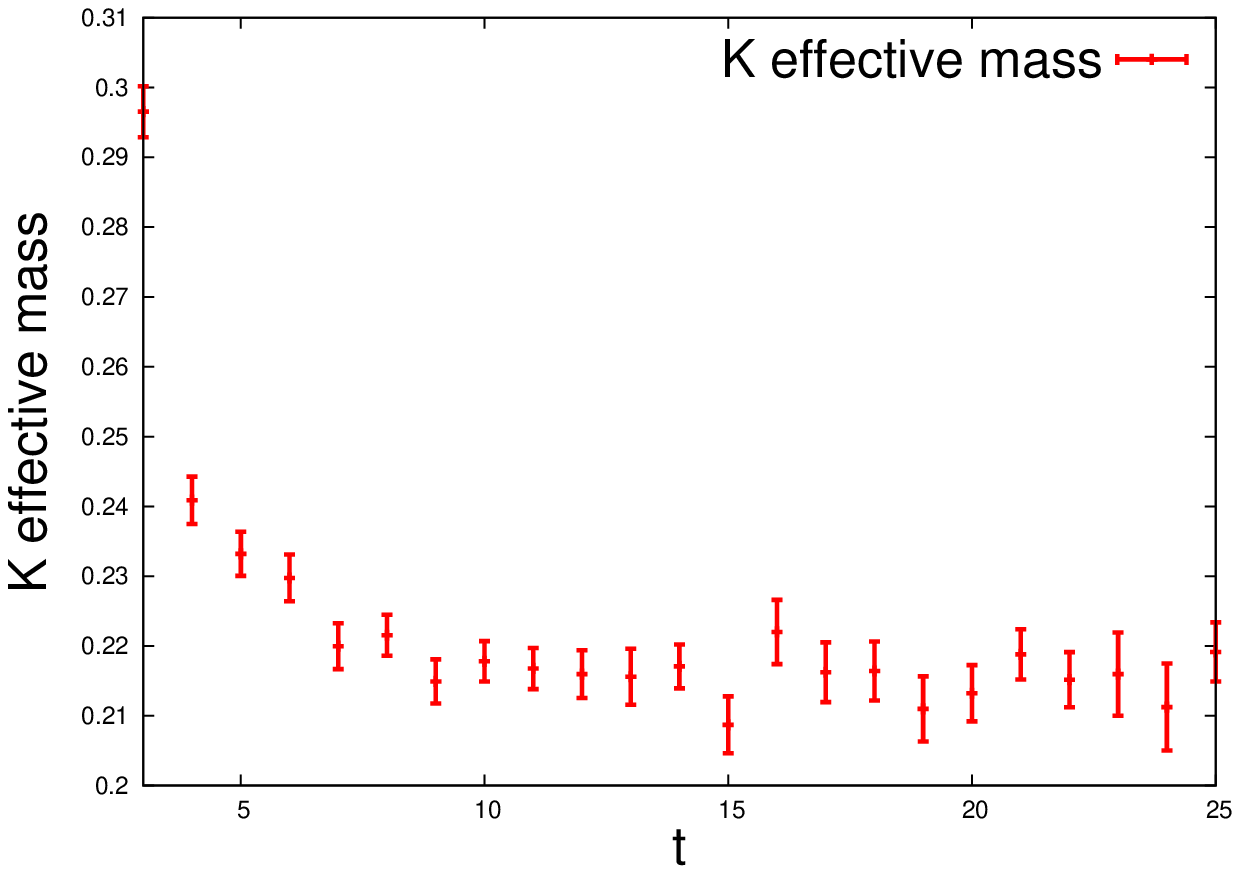} \\
		\end{tabular}
		\caption{Effective mass plots for the kaon correlation functions on the $48^3$ ensemble (left) and $64^3$ ensemble (right).}
		\label{fig:keffm}
		\end{center}
	\end{figure}
\begin{figure}[t]
		\begin{center}
		\begin{tabular}{ccc}
		\includegraphics[width=0.445\hsize]{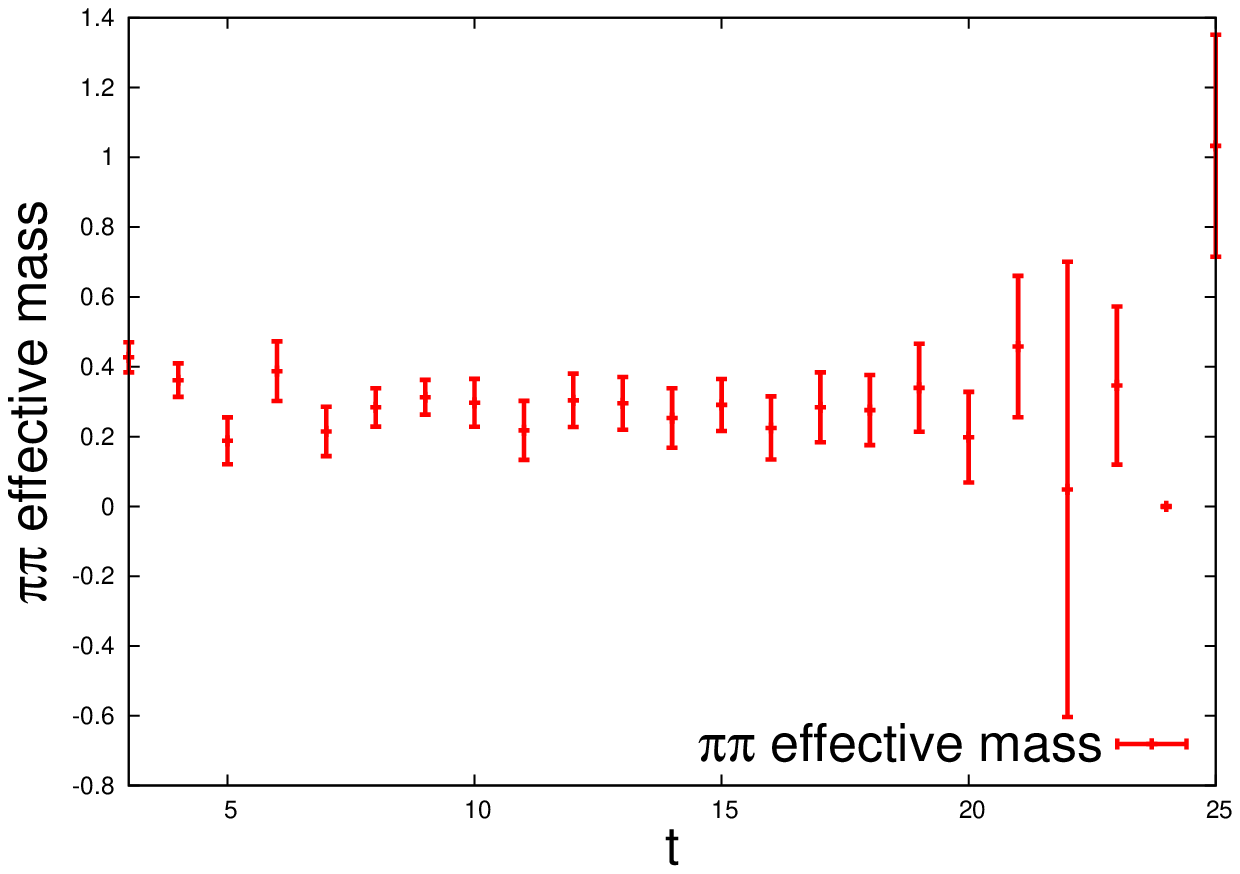} \qquad \includegraphics[width=0.445\hsize]{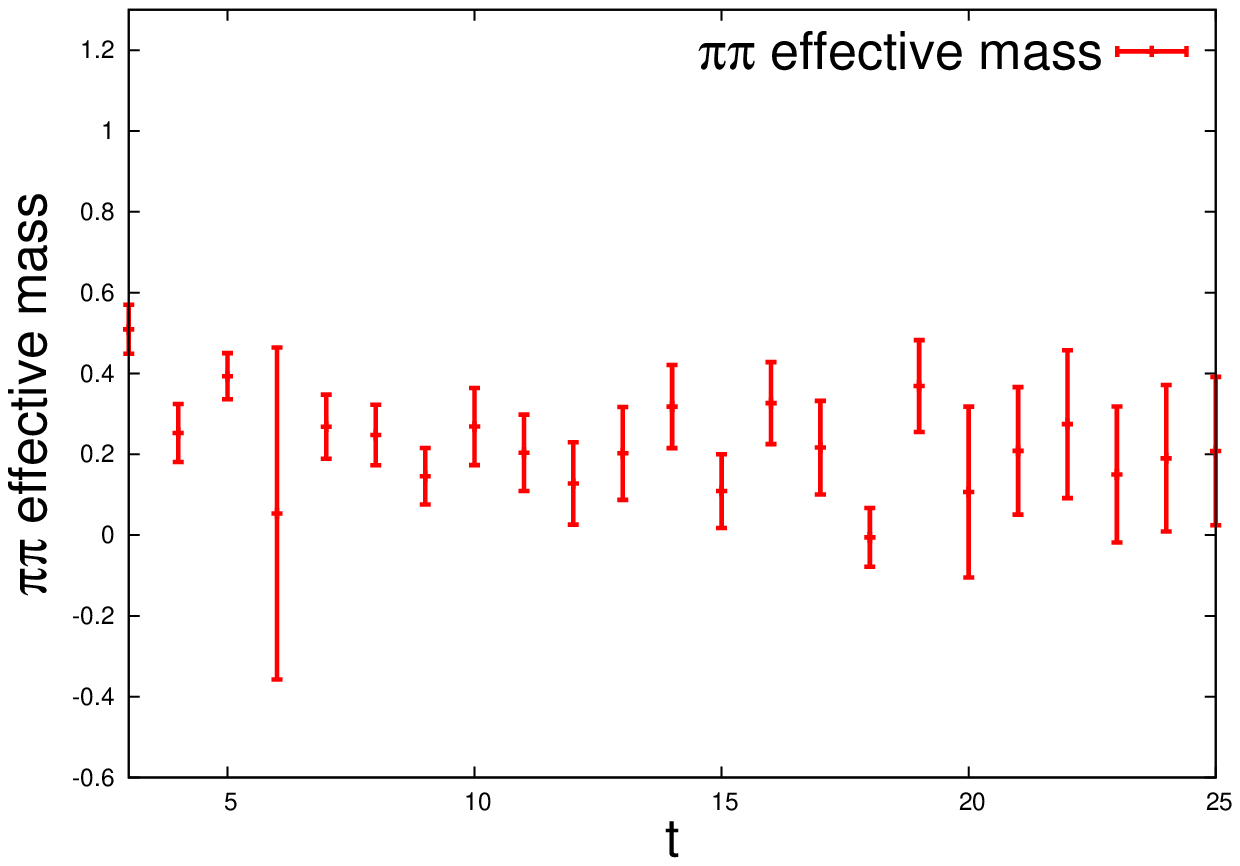} \\
		\end{tabular}
		\caption{Effective mass plots for the two-pion correlation functions on the $48^3$ ensemble (left) and $64^3$ ensemble (right).}
		\label{fig:pipieffm}
		\end{center}
	\end{figure}

The pion ($m_\pi$) and kaon masses ($m_K$) as well as the energies of the $I=2$ two-pion \rm{stat}e ($E_{\pi\pi}$) obtained on the two ensembles are shown in Table \,\ref{tab:masse}. The fitting ranges used for pion and kaon masses as well as two pion energies were from 10 to 86 on the $48^3$ ensemble and from 10 to 118 on the $64^3$ ensemble. These choices were motivated by the plateaus in the effective mass plots shown in Figs. \ref{fig:keffm}\,-\,\ref{fig:pipieffm}. 
The effective mass of the kaon, $m_K^\textrm{eff}$, is defined numerically by 
 the ratio
\begin{equation}
	\frac{C_{K}(t+1)}{C_{K}(t)} = \frac{\mathrm{cosh}(m_K^\textrm{eff}(t+1-T/2))}{\mathrm{cosh}(m_K^\textrm{eff}(t-T/2))},
\end{equation} 
and the two-pion effective mass, $E^\textrm{eff}_{\pi\pi}$, is found by inverting
\begin{equation}\label{eq:effmasspipi}\textrm{\textrm{}}
	\frac{C_{\pi\pi}(t+2)-C_{\pi\pi}(t+1)}{C_{\pi\pi}(t+1)-C_{\pi\pi}(t)} = \frac{e^{-E^\textrm{eff}_{\pi\pi}(t+2)} + e^{-E^\textrm{eff}_{\pi\pi}(T-t-2)} - e^{-E^\textrm{eff}_{\pi\pi}(t+1)} + e^{-E^\textrm{eff}_{\pi\pi}(T-t-1)}}{e^{-E^\textrm{eff}_{\pi\pi}(t+1)} + e^{-E^\textrm{eff}_{\pi\pi}(T-t-1)} - e^{-E^\textrm{eff}_{\pi\pi}t} + e^{-E^\textrm{eff}_{\pi\pi}(T-t)}}.
\end{equation}
The two-point correlation functions $C_K$ and $C_{\pi\pi}$  are defined explicitly in Eq.\,(\ref{eq:2pt}) below
and the differences in the numerator and denominator on the left-hand side of Eq.\,(\ref{eq:effmasspipi}) are introduced to eliminate the constant $C$ in Eq.\,(\ref{eq:cpipi}).


The pion and kaon masses correspond closely to their physical values. We will explain below that the pions are given a momentum $\pi/L$ in each of the three spatial directions and from the table we see that with this choice $E_{\pi\pi}\simeq m_K$ and the 
$K\rightarrow\pi\pi$ matrix elements correspond to the on-shell (within statistical errors) decay of a kaon in the center-of-mass frame. We now discuss the evaluation of the matrix elements.

\section{Evaluation of the bare matrix elements}\label{sec:bare}
		$K\rightarrow \pi\pi$ decay amplitudes are defined by
		\begin{equation}
		\sqrt{2}\,A_{2,0}\,e^{i\delta _{2,0}} = \langle (\pi\pi)_{I=2,0} \mid H_W \mid K^0\rangle,
		\end{equation}
		where $H_W$ is the component of the weak Hamiltonian which changes the strangeness by one unit.
		The weak Hamiltonian can be separated into short and long distance contributions by using the operator product expansion:
		\begin{equation}
		H_W = \frac{G_F}{\sqrt{2}}V_{ud}^*V_{us}\,\sum_{i}C_i (\mu) Q_i(\mu),
		\label{eq:hw}
		\end{equation}
		where $G_F$ is the Fermi constant, $V_{us}$ and $V_{ud}$ are Cabibbo-Kobayashi-Maskawa (CKM) matrix elements, the $Q_i$ are all the possible dimension-6 operators which contribute to the decay and $C_i$ are the corresponding Wilson coefficients which contain information about the short distance physics. The $C_i$ take the form $C_i=z_i+\tau y_i$ where $\tau$ is the ratio of CKM matrix coefficients $\tau = -\frac{V_{ts}^*V_{td}}{V_{us}^*V_{ud}}$.
		
		In this paper we only consider $\Delta I = 3/2$ decays where the two-pion final \rm{stat}e has total isospin 2. The nonperturbative contribution to the decay amplitude is contained in the matrix elements:
\begin{equation}
M_i^{K^0} \equiv \langle (\pi\pi)^{I=2}_{I_3=0} \mid Q^{\Delta I = 3/2}_{\Delta I_3 = 1/2,i} \mid K^0\, \rangle\quad\textrm{and}\quad
		M_i^{K^+} \equiv \langle (\pi\pi)^{I=2}_{I_3=1} \mid Q^{\Delta I = 3/2}_{\Delta I_3 = 1/2,i} \mid K^+ \,\rangle.\label{eq:mikdef}
\end{equation}
		There are only three operators which contribute to $A_2$, which we label according to their chiral $SU(3)_L\times SU(3)_R$ transformation properties. We have one (27,1) operator and two electroweak penguin operators labeled (8,8) and $(8,8)_{\textrm{mx}}$, where the subscript {\footnotesize $\textrm{mx}$} denotes a color mixed operator. 
		Explicitly, the operators are given by
		\begin{eqnarray}
			Q^{\Delta I = 3/2}_{(27,1)} &=& (\bar s_i d_i)_L\left( \bar u_j u_j - \bar d_jd_j\right)_L + (\bar s_i u_i)_L(\bar u_jd_j)_L,\label{eq:Q271}\\
			Q^{\Delta I = 3/2}_{(8,8)} &=& (\bar s_i d_i)_L\left( \bar u_j u_j - \bar d_jd_j\right)_R + (\bar s_i u_i)_L(\bar u_jd_j)_R,\label{eq:Q88}\\
			Q^{\Delta I = 3/2}_{(8,8)\textrm{mx}} &=& (\bar s_i d_j)_L\left( \bar u_j u_i - \bar d_jd_i\right)_R + (\bar s_i u_j)_L(\bar u_jd_i)_R.\label{eq:Q88mx}
		\end{eqnarray}
		The subscripts L and R denote the left- and right-handed spin structures respectively:
		\begin{equation}
			(\bar q_1 q_2)_L = \bar q_1 \gamma^\mu (1 - \gamma^5) q_2 \quad\mathrm{and}\quad
			(\bar q_1 q_2)_R = \bar q_1 \gamma^\mu (1 + \gamma^5) q_2.
		\end{equation}
		The Lorentz indices are understood to be contracted between the two parentheses in each of the operators in Eqs.\,(\ref{eq:Q271})\,-\,(\ref{eq:Q88mx}) and $i,j$ are color indices which are summed from 1 to 3.
		
		Below we will confirm the feature found in our earlier work\,\cite{Blum:2011ng,Blum:2012uk}
 that the dominant contribution to Re($A_2$) comes from the (27,1) operator, while the dominant contribution to Im($A_2$) in the $\overline{\mathrm{MS}}$ scheme at 3\,GeV comes from the $(8,8)_{\textrm{mx}}$ operator.
		We can now write the expressions for the $A_2$ amplitude, which are
\begin{equation}
			A_2 =\frac{G_F}{\sqrt{2}}V_{ud}^*V_{us}\,\sum_{i}C_i (\mu) \left( \frac{1}{\sqrt 2} M_i^{K^0} \right)
			=\frac{G_F}{\sqrt{2}}V_{ud}^*V_{us}\,\sum_{i}C_i (\mu) \left( \frac{1}{\sqrt 3} M_i^{K^+} \right).
\end{equation}		
The relative factor between the two expressions is due to the different Clebsch-Gordan coefficients.
		
	A major challenge in the calculation of $A_2$ (and even more so in the calculation of $A_0$) is to ensure that the   
pions have physical momenta. In the center-of-mass frame with periodic boundary conditions, the ground \rm{stat}e for the two-pion system has each pion at rest. The evaluation of matrix elements at physical kinematics therefore corresponds to  the contribution from an excited two-pion \rm{stat}e resulting in a considerable loss of precision.  
We can avoid the necessity of multiexponential fits to extract the excited state contribution by utilizing the technique suggested in~\cite{Kim:2003xt,Kim:2005gka}  and applied successfully in our original calculation of A2~\cite{Blum:2011ng,Blum:2012uk}: we introduce antiperiodic boundary conditions for the (valence) d-quark in all three spatial directions, and periodic boundary conditions for the u- and s-quarks~\cite{Kim:2003xt}. 
We then exploit the Wigner-Eckart theorem to relate $K^+\rightarrow\pi^+\pi^0$ matrix elements to those for the unphysical transition 
$K^+\rightarrow\pi^+\pi^+$. The relation is 
	\begin{equation}\label{eq:WE}
		\underbrace{\langle (\pi\pi)^{I=2}_{I_3=1}\mid}_{\frac{1}{\sqrt{2}}(\langle \pi^+\pi^0 \mid + \langle \pi^0\pi^+ \mid)} Q^{\Delta I = 3/2}_{\Delta I_3 = 1/2,i} \mid K^+ \rangle = \frac{3}{2} \underbrace{\langle (\pi\pi)^{I=2}_{I_3=2}\mid}_{\langle \pi^+\pi^+ \mid} Q^{\Delta I = 3/2}_{\Delta I_3 = 3/2,i} \mid K^+ \rangle\:.
	\end{equation}
The indices $I$ and $I_3$ label the two-pion state's total and third component of isospin respectively. 
With antiperiodic boundary conditions in three spatial directions, the $|\pi^+\pi^+\rangle$ ground \rm{stat}e has total momentum $\vec{0}$, with each pion having momentum $|\vec{p}_\pi|=\sqrt{3}\,\pi/L$. It can be seen from Table~\ref{tab:masse} that $E_{\pi\pi}$ is very close to $m_K$ on both the $64^3$ and $48^3$ ensembles. 
(For the smaller physical volume in our original calculation~\cite{Blum:2011ng,Blum:2012uk}, we imposed antiperiodic boundary conditions for the $d$-quark in two spatial directions in order to achieve $E_{\pi\pi}\simeq m_K$.) Note that with both periodic and antiperiodic boundary conditions on the $d$-quark, the lowest momentum of the $\pi^0$ meson is zero; this motivates the use of the Wigner-Eckart theorem to reformulate the calculation to that of a matrix element with a $|\pi^+\pi^+\rangle$ final \rm{stat}e. 

The operators $Q^{\Delta I = 3/2}_{\Delta I_3 = 3/2}$ which appear on the right-hand side of Eq.\,(\ref{eq:WE}), and which correspond to the $Q^{\Delta I = 3/2}_{\Delta I_3 = 1/2}$ operators in Eqs.\,(\ref{eq:Q271})\,-\,(\ref{eq:Q88mx}), are
\begin{eqnarray}
\label{eq:4qop}
	Q_{(27,1)} &=& (\bar s_i d_i)_L (\bar u_j d_j)_L,\quad
	Q_{(8,8)} = (\bar s_i d_i)_L (\bar u_j d_j)_R,\quad
	Q_{(8,8)\textrm{mx}} = (\bar s_i d_j)_L (\bar u_j d_i)_R.
\end{eqnarray}
To simplify the notation we have dropped the labels $\Delta I=3/2$ and $\Delta I_z=3/2$ on the operators in Eq.\,(\ref{eq:4qop}); this will be implicit in the following. In this paper we compute the $K\to\pi\pi$ matrix elements of the three operators in Eq.\,(\ref{eq:4qop}).

The factor of $3/2$ in Eq.\,(\ref{eq:WE}) is a combination of $\sqrt{3}/2$ coming from the Clebsch-Gordan coefficients and the Wigner-Eckart theorem, and a further $\sqrt{3}$ corresponding to the simple choice for the normalization of operators in Eq.\,(\ref{eq:4qop}). 
The amplitude $A_2$ is given in terms of the $K^+\rightarrow \pi^+\pi^+$ matrix elements $M_i$ by
\begin{equation}
	A_2 =\frac{G_F}{\sqrt{2}}V_{ud}^*V_{us}\,\frac{\sqrt{3}}{2}\,\sum_{i}C_i (\mu)  M_i.
\end{equation}
Since it is the $K^+\rightarrow \pi^+\pi^+$ matrix elements which we compute directly in this paper, we choose the compact notation $M_i\equiv M_i^{K^+\to\pi^+\pi^+}$. The label $i$ runs over the three operators in Eq.\,(\ref{eq:4qop}).

\subsection{Evaluation of the correlation functions} 

The bare matrix elements are obtained from the computation of two- and three-point correlation functions. The three-point functions are
		\begin{equation}
			C^{K\rightarrow \pi\pi}_i (t_{\mathrm{op}}) = \langle 0 \mid \sigma_{\pi\pi}(t_{\pi\pi})\,Q_i(t_{\mathrm{op}})\, \sigma^\dagger_K(0) \mid 0 \rangle,
		\label{eq:3ptfn}
		\end{equation}
where $Q_i$ is one of the three operators in Eq.\,(\ref{eq:4qop}) and $\sigma_K$ and $\sigma_{\pi\pi}$ are interpolating operators for the kaon and two-pion \rm{stat}e respectively. For $\sigma_K$ and $\sigma_{\pi\pi}$ we take
Coulomb gauge-fixed wall-source operators defined as follows:
		\begin{eqnarray}\label{eq:sigmaK}
			\sigma_K(t) &\equiv & \sum_{\vec x_1, \vec x_2} \bar{s}(\vec x_1, t)\, \gamma^5 \,u(\vec x_2, t),\\
			\sigma_{\pi\pi}(t) &\equiv & \left[\bar d(t) \gamma^5 u(t)\right] \left[\bar d(t) \gamma^5 u(t)\right],
			\label{eq:sigmapipi}
		\end{eqnarray}
		where in (\ref{eq:sigmapipi}) we have used the cosine momentum sources for the $d$-quark:
		\begin{equation}\label{eq:cossource}
			d(t) = \sum_{x,y,z} d(x,y,z,t)\cos(xp_x)\cos(yp_y)\cos(zp_z)\,.
		\end{equation}
$d(x,y,z,t)$ represents the $d$-quark field and the components of momenta satisfy $p_x=p_y=p_z=\pi/L$. 
Just as for the $u$-quark source in Eq.\,(\ref{eq:sigmaK}), the $u$-quark sources in $\sigma_{\pi\pi}$ shown in Eq.\,(\ref{eq:sigmapipi})
are given zero momentum by summing them over the full spatial volume, evaluated in the Coulomb gauge.
As explained in Ref.~\cite{Blum:2012uk} the cosine
source described above creates $d$-quarks with both
signs for each component of the three momentum 
$\pm p_i$, for $i=x$, $y$ and $z$. This will then
produce pairs of pions with total momentum in each
direction of $\pm 2\pi/L$ in addition to the desired
value of $\vec 0$.  For the three-point functions
described in Eq.\,(\ref{eq:3ptfn}), the zero total momentum of
the decaying kaon and three-momentum conservation
imply that the nonzero $\pi$-$\pi$ momenta cannot
occur.  For the two-point function defined in Eq.\,(\ref{eq:2pt})
below we use a $\pi$-$\pi$ sink which is different
from the source and which explicitly projects onto
$\pi$-$\pi$ states with zero total momentum, as
described in Ref.~\cite{Blum:2012uk}.   A further
subtlety, not described in that reference, relates
to the possible angular momentum of the two-pion
state.  For our two identical $\pi^+$ bosons which
carry equal but opposite momenta, there are actually
four possible states given our boundary conditions.
Specifically, the $\pi^+$ which carries $p_x = +\pi/L$
may have four possible values for the other momentum
components: $p_y = \pm\pi/L$ and $p_z=\pm\pi/L$.
These four states form a four-dimensional representation of the cubic symmetry group, which decomposes into two irreducible representations: a singlet ($A_1$) and a triplet ($T_2$), out of which only $A_1$ contains an s-wave contribution. Since the lowest energy level of the finite-volume $I=2$ s-wave spectrum of the $A_1$ representation is nearly degenerate with the lowest energy level of the d-wave spectrum of the $T_2$ representation, it is important that we use the cubically symmetrical source specified in Eq.\,(\ref{eq:cossource}) which couples only to the $A_1$ state of interest.

\begin{figure}
		\begin{center}
		\includegraphics[scale=0.6]{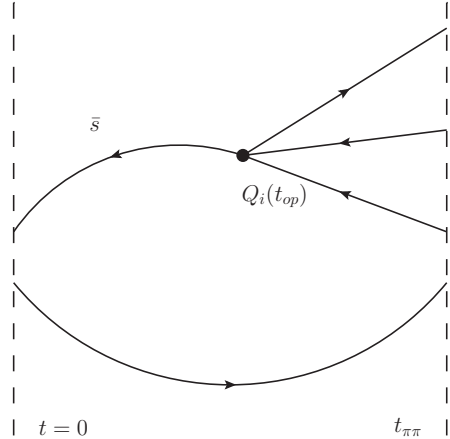}
		\caption{Diagrammatic representation of the $K \rightarrow \pi\pi$ three-point function defined in Eq.\,(\ref{eq:3ptfn}). The strange-quark propagator is explicitly labeled, the remaining lines represent light-quark propagators.}
		\label{fig:3ptfn}
		\end{center}
	\end{figure}

The spinor and color labels are contracted within each set of square parentheses in Eq.\,(\ref{eq:sigmapipi}). 
A schematic diagram of the correlation function $C^{K\rightarrow \pi\pi}_i (t_{\mathrm{op}})$ is shown in Fig.\,\ref{fig:3ptfn}.

\begin{figure}[t]
		\begin{center}
		\includegraphics[width=0.45\hsize]{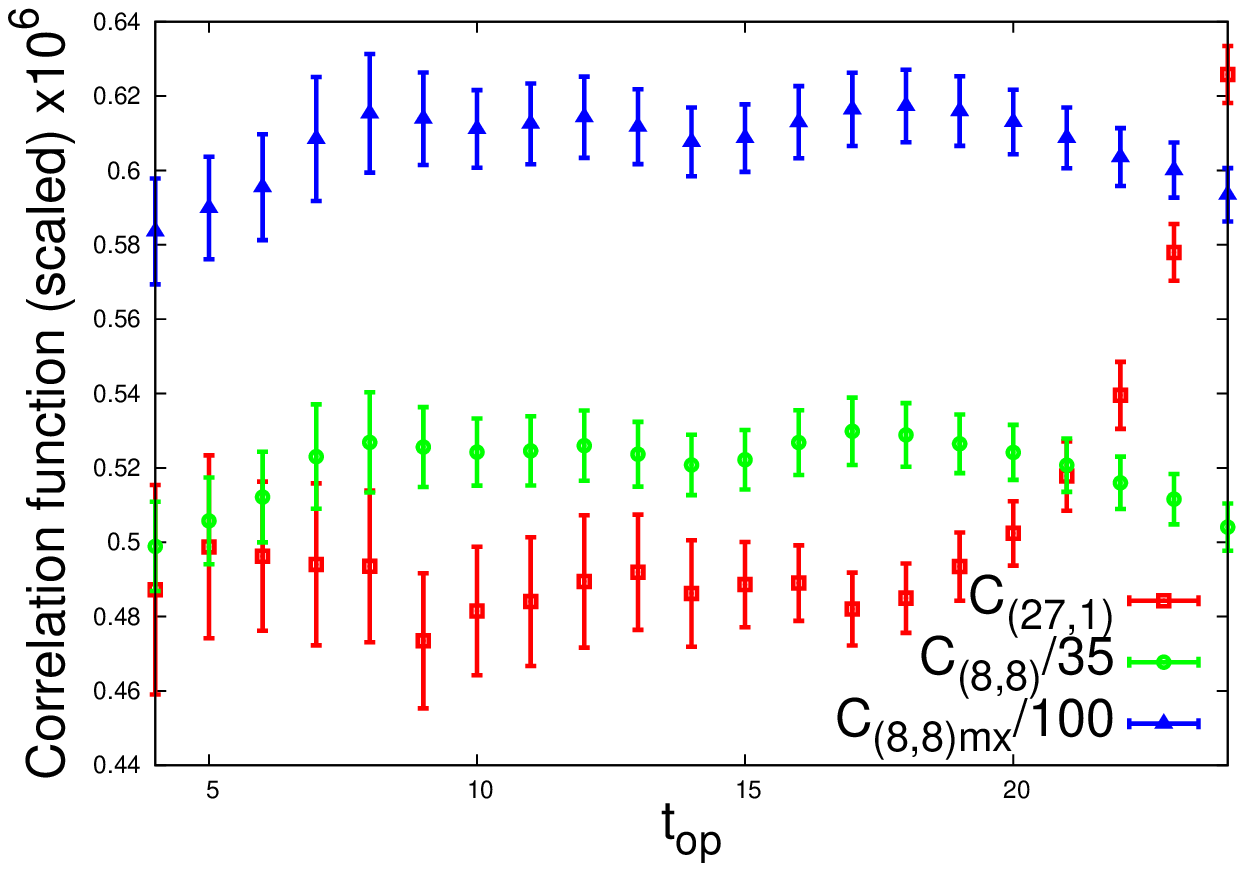} \qquad \includegraphics[width=0.45\hsize]{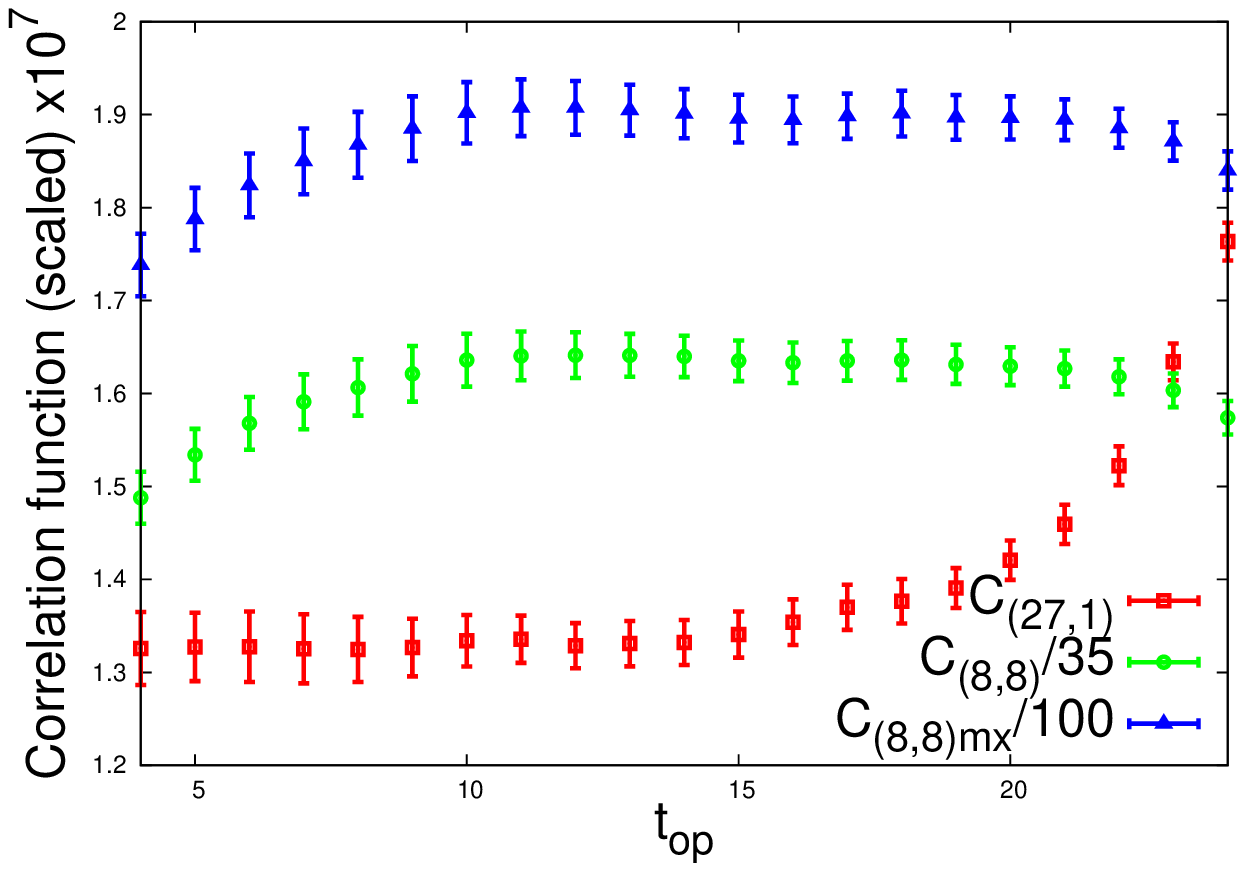} \\
		\caption{$K\rightarrow\pi\pi$ three-point correlation function on the $48^3$ lattice (left) and $64^3$ lattice (right) with a kaon-pion separation of $t_{\pi\pi} = 26$.}
		\label{fig:3pt48cube}
		\end{center}
	\end{figure}

We have evaluated $C^{K\rightarrow \pi\pi}_i (t_{\mathrm{op}})$ for a range of values of the source-sink separations $t_{\pi\pi}$. For the $48^3$ ($64^3$) ensemble
we performed the calculations for values of $t_{\pi\pi}$ between 24 and 39 (26 and 36).
These separations were chosen to be large enough for the plateau region to give a reliable fit and small enough for the around-the-world effects to be small.
The fitting ranges were chosen to be from 10 to $t_{\pi\pi}-10$ for both ensembles. These choices are motivated by the locations of plateau regions in Fig.\,\ref{fig:3pt48cube}.

For sufficiently large time separations $t_\mathrm{op}$ and $t_{\pi\pi}-t_\mathrm{op}$, the expected time dependence of $C^{K\rightarrow \pi\pi}_i(t_{\mathrm{op}})$ is
	\begin{equation}
		C^{K\rightarrow \pi\pi}_i(t_{\mathrm{op}}) = N_{\pi\pi}\,N_{K}\,M^\mathrm{bare}_{i}\,e^{-(m_K-E_{\pi\pi})t_{\mathrm{op}}}\,e^{-E_{\pi\pi}t_{\pi\pi}}\,,
		\label{eq:3ptfunctiontimedependence}
	\end{equation}	
where 
\begin{equation}
		N_{\pi\pi} = \left| \langle \pi\pi\, |\, \sigma_{\pi\pi}(0)\, |\, 0 \rangle\right|\quad\mathrm{and}\quad
		N_K = \left| \langle K | \,\sigma_K(0)\, | 0 \rangle \right|.
	\end{equation}
We have introduced the label ``bare" as a reminder that $M^\mathrm{bare}_{i}$ are matrix elements of the bare operators in the lattice regularization which we are using. The renormalization of the operators is discussed in the following section. 
For illustration, in Fig.\,\ref{fig:3pt48cube} we plot $C^{K\rightarrow \pi\pi}_i(t_{\mathrm{op}})$ computed on each of the two ensembles for $t_{\pi\pi}=26$. The observed plateaus are a manife\rm{stat}ion of the fact that the volumes have been tuned so that $E_{\pi\pi}\simeq m_K$ [cf. Eq. (\ref{eq:3ptfunctiontimedependence})].
	
We obtain the matrix elements $M_i$ by fitting Eq.\,(\ref{eq:3ptfunctiontimedependence}), using the values of $N_{\pi\pi}$, $N_K$, $m_K$ and $E_{\pi\pi}$ obtained from fitting (under the jackknife) the correlation functions,
	\begin{equation}
		C_{\pi\pi}(t) = \langle 0\,| \sigma^\dagger_{\pi\pi}(t,\vec{p}=0) \,\sigma_{\pi\pi}(0)|\, 0 \rangle\quad\mathrm{and}\quad
		C_K(t) = \langle 0 | \sigma_K(t) \sigma^\dagger_K(0) | 0 \rangle,\label{eq:2pt}
	\end{equation}
which have the following time dependence:
	\begin{eqnarray}
		C_{\pi\pi}(t) &\xrightarrow[t\rightarrow \infty]{}& \left| N_{\pi\pi} \right|^2 \left( e^{-E_{\pi\pi}} + e^{-E_{\pi\pi}(T-t)} + C\right),\label{eq:cpipi}\\
		C_K(t) &\xrightarrow[t\rightarrow \infty]{}& \left| N_K \right|^2 \left( e^{-m_Kt} + e^{-m_K(T-t)}\right).
	\end{eqnarray}
	The ``$t\rightarrow\infty$'' limit should be understood as taking a sufficiently large time separation so that excited \rm{stat}e contributions are negligible. Introducing the constant $C$ in Eq.\,(\ref{eq:cpipi}) allows one to account for possible around-the-world effects in $C_{\pi\pi}$.

\begin{figure}[t]
		\begin{center}
		\begin{tabular}{ccc}
		\includegraphics[width=0.445\hsize]{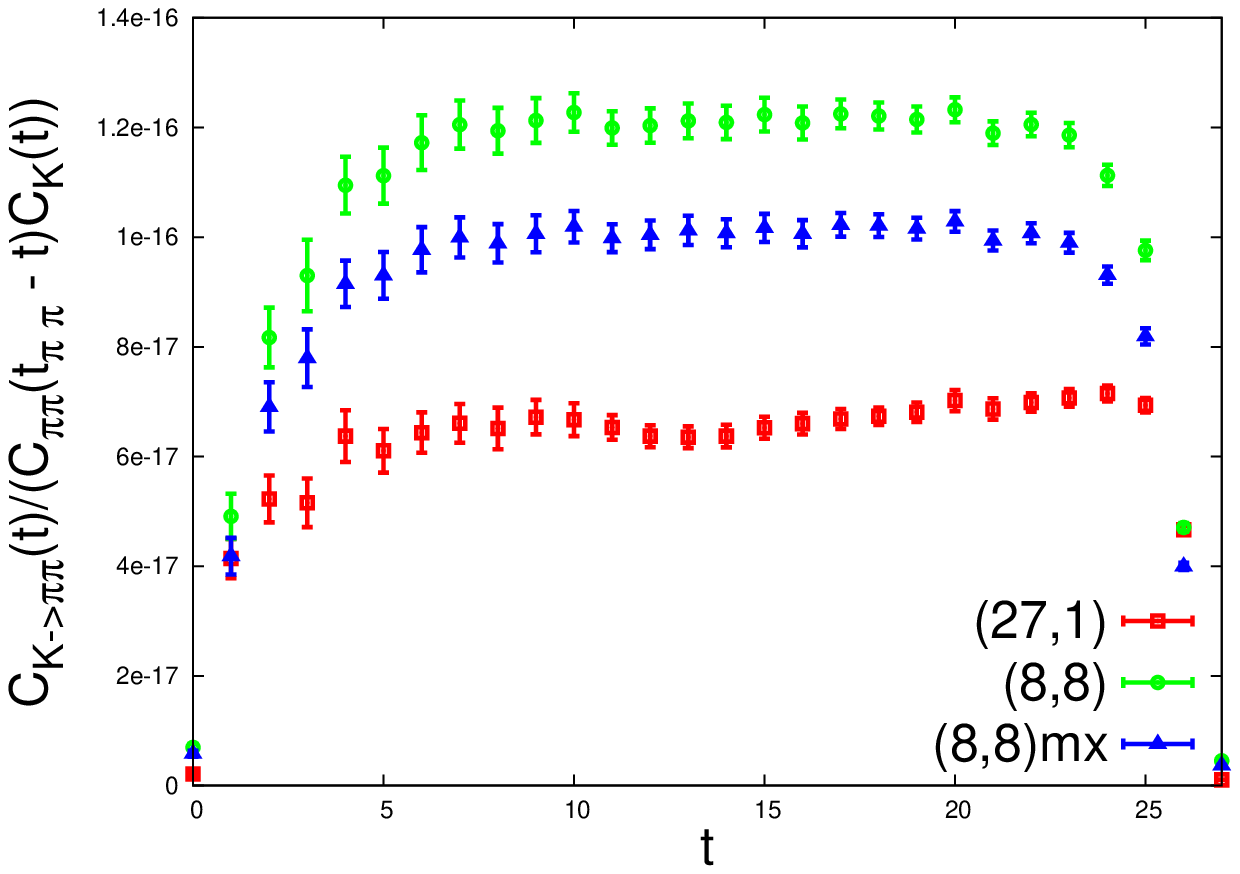} \qquad \includegraphics[width=0.445\hsize]{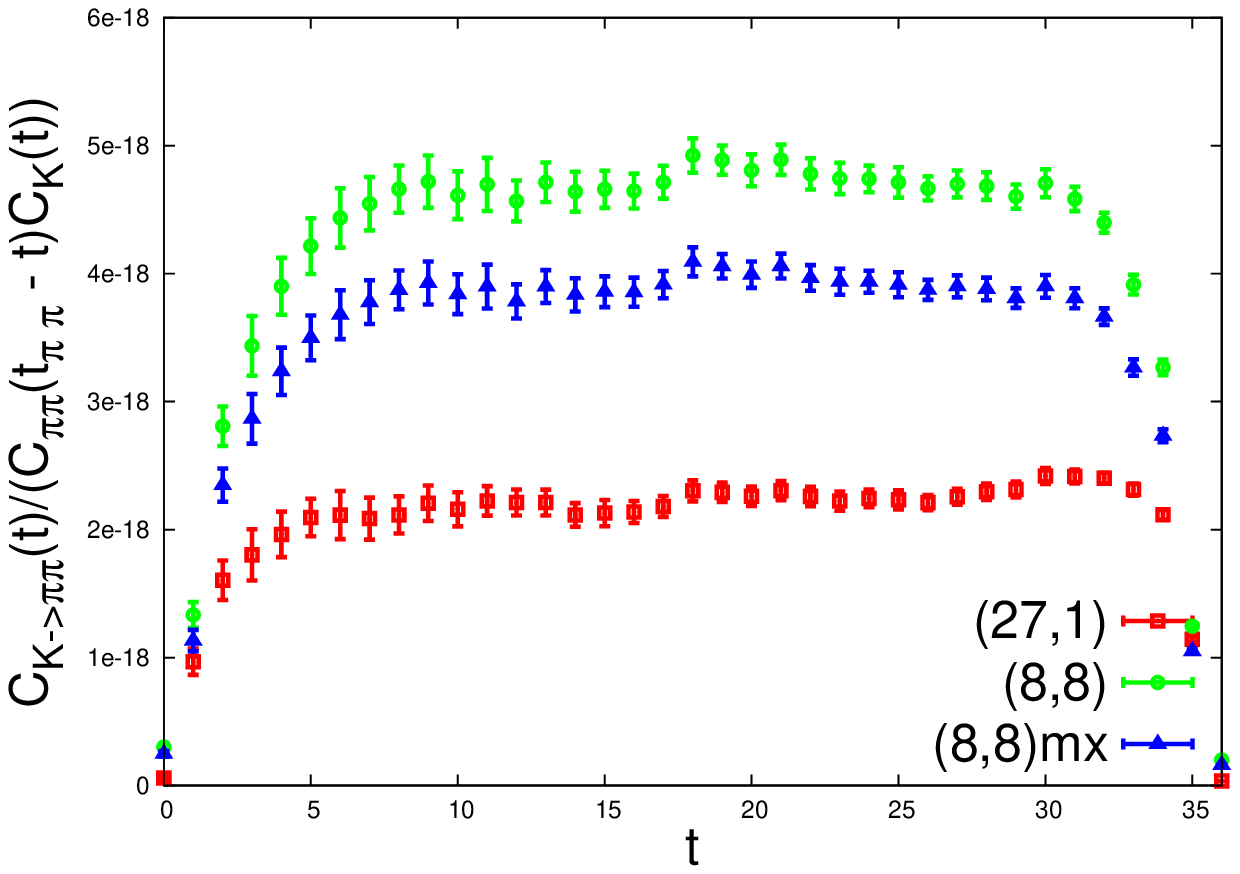} \\
		\end{tabular}
		\caption{Ratios of $K\rightarrow\pi\pi$ three-point correlation function to the two-point functions (Eq.\,(\ref{eq:3pt2ptratio})) on the $48^3$ lattice (left) and the $64^3$ lattice (right) with a kaon-pion separations of $t_{\pi\pi} = 27$ and $36$ respectively.}
		\label{fig:3ptratios}
		\end{center}
	\end{figure}

As a check, we can also construct the time-independent ratio of the correlation functions:
	\begin{equation}
		\frac{C^i_{K\rightarrow \pi\pi}(t)}{C_K(t) \,C_{\pi\pi}(t_{\pi\pi} - t)} = \frac{M^\mathrm{bare}_i}{N_{\pi\pi}N_K}.
		\label{eq:3pt2ptratio}
	\end{equation}
	This ratio is plotted for $t_{\pi\pi}=26$ in Fig.\,\ref{fig:3ptratios}.  As anticipated, all three operators exhibit a constant behavior in the region where the contribution from excited \rm{stat}es is negligible.
	Equation~(\ref{eq:3pt2ptratio}) is expected to hold in the region $0\ll t\ll t_{\pi\pi}\ll T$, where $T$ is the total time extent of the lattice. In this region ``around-the-world'' effects arising from different time orderings of the operators can be neglected.

The values of the bare $K^+\rightarrow \pi^+\pi^+$ matrix elements are shown in Table~\ref{tab:mi}. The entries have been obtained by performing weighted averages (under the jackknife) over the values obtained for each choice of $t_{\pi\pi}$.

	\begin{table}[t]
	\begin{center}
	\begin{tabular}{|c|c|c|c|}
	\hline
	& $a^3M_{(27,1)}^{\mathrm{bare}}$ & $a^3M_{(8,8)}^{\mathrm{bare}}$ & $a^3M_{(8,8)\textrm{mx}}^{\mathrm{bare}}$\\
	\hline
	$48^3$ ensemble & $3.700(35) \times 10^{-4}$ &$9.171(69) \times 10^{-3}$ &$3.058(23) \times 10^{-2}$ \\
	$64^3$ ensemble  & $1.371(11) \times 10^{-4}$& $3.942(39) \times 10^{-3}$& $1.308(13) \times 10^{-2}$\\
	\hline
	\end{tabular}
	\end{center}
	\caption{Results for the bare $K^+\rightarrow \pi^+\pi^+$ matrix elements in lattice units. Only \rm{stat}istical errors are shown.}
	\label{tab:mi}
	\end{table}
	
\section{Renormalization of the Operators}\label{sec:NPR}
	
	Having determined the matrix elements of the bare operators in the lattice regularization we now have to combine them with the remaining factors in Eq.\,(\ref{eq:hw}) to obtain $A_2$. The Wilson coefficients [$C_i(\mu)$] and composite operators [$Q_i(\mu)$] appearing in Eq.\,(\ref{eq:hw}) are separately renormalization scheme and scale ($\mu$) dependent. To obtain the physical amplitudes they must be combined in the same scheme and at the same scale. The $C_i(\mu)$ are calculated in perturbation theory for which it is convenient to use the \MSbar-NDR scheme (called \MSbar in the following). NDR stands for ``naive dimensional regularization'' prescription for the $\gamma^5$ matrix, which preserves the anticommutation relations with other gamma matrices \cite{Buchalla:1995vs}. The matrix elements calculated in Sec.\,\ref{sec:bare}, on the other hand, were obtained using bare operators with the lattice spacing as the ultraviolet regulator with the lattice discretization of QCD. 
The operators can be renormalized nonperturbatively, but only into schemes for which the renormalization condition can be imposed on lattice Green's functions. 
The \MSbar scheme, which is based on dimensional regularization cannot be simulated in a lattice computation. Our procedure is to start by renormalizing the operators non-perturbatively into schemes which can be simulated, specifically the ``regularization-independent symmetric momentum'' (RI-SMOM) schemes~\cite{Sturm:2009kb} as described in detail in \cite{Blum:2012uk}  and briefly summarized below. The matching between the RI-SMOM and \MSbar schemes is necessarily performed in perturbation theory and is currently known at one-loop order. (Below we also present the matrix elements in two RI-SMOM schemes so that if the perturbative coefficients are calculated to higher order in the future, these matrix elements can be used to reduce the systematic uncertainty in $A_2$ due to the truncation of the perturbation series.) 

We now briefly summarize the renormalization procedure. We write the five-point 
amputated Green's functions of the three operators in Eq.\,(\ref{eq:4qop}) as a three-component vector 
$\Lambda=(\Lambda_{(27,1)},\Lambda_{(8,8)},\Lambda_{(8,8)\textrm{mx}})\equiv(\Lambda_1,\Lambda_2,\Lambda_3)$, and impose a renormalization condition of the form
\begin{equation}
\label{eq:renorm_cond}
  P\left\{\Lambda^{R}(\mu)\right\} = F \;,
\end{equation}
where $P$ is a vector of projectors and $F$ the corresponding tree-level matrix. Denoting the tree-level contribution by the superscript $(0)$ and including explicitly the spinor and color labels, the matrix $F$ is given by 
\begin{equation}
P_i\left\{\Lambda_j^{(0)}\right\} \equiv
\left[ P_i \right]_{\beta \alpha ; \delta \gamma}^{BA;DC} 
\left[\Lambda_j^{(0)}\right]_{\alpha \beta ; \gamma \delta}^{AB;CD} = F_{ij}.\,
\end{equation} 
Here greek letters label spinor components, the uppercase roman letters 
represent color indices and $i,j=1,2,3$ denote the operators and projectors. 
For illustration, the tree-level value of the Green's function of $Q^{(27,1)}$ is
\begin{align}
\left[\Lambda_1^{(0)}\right]_{\alpha \beta ; \gamma \delta}^{AB;CD}
= &
\left[(\gamma^\mu)_{\alpha\beta} (\gamma^\mu)_{\gamma\delta} + (\gamma^\mu\gamma_5)_{\alpha\beta} (\gamma^\mu\gamma_5)_{\gamma\delta} \right]
\delta^{AB}\delta^{CD} \nonumber\\
 &\quad-
\left[(\gamma^\mu)_{\alpha\delta} (\gamma^\mu)_{\gamma\beta} + (\gamma^\mu\gamma_5)_{\alpha\delta} (\gamma^\mu\gamma_5)_{\gamma\beta} \right]
\delta^{AD}\delta^{BC} \;.
\end{align}
For the renormalization we only consider the parity-even component of the four-quark operators.

The choice of projectors is not unique and  
we implement two different sets  
known as the $\gamma_\mu$ and $\s{q}$-projectors, given explicitly by
\begin{equation}
  \left[ P^{(\gamma^\mu)}\right]_{\beta\alpha ; \delta \gamma}^{JI;LK}
  = \left(
  \begin{array}{l}
    \left[ (\gamma^\mu)_{\beta\alpha} (\gamma^\mu)_{\delta\gamma}
      + (\gamma^\mu \gamma^5)_{\beta\alpha}(\gamma^\mu\gamma^5)_{\delta\gamma} \right]
    \delta^{JI}\delta^{LK}\\
    \left[ (\gamma^\mu)_{\beta\alpha} (\gamma^\mu)_{\delta\gamma} -
      (\gamma^\mu \gamma^5)_{\beta\alpha}(\gamma^\mu\gamma^5)_{\delta\gamma} \right]
    \delta^{JI}\delta^{LK}\\
    \left[ (\gamma^\mu)_{\beta\gamma} (\gamma^\mu)_{\delta\alpha} -
      (\gamma^\mu \gamma^5)_{\beta\gamma}(\gamma^\mu\gamma^5)_{\delta\alpha} \right]
    \delta^{JK}\delta^{LI}
  \end{array}
  \right) 
\end{equation}
and
\begin{equation}
  \left[ P^{(\s{q})} \right]_{\beta \alpha \beta ; \delta \gamma}^{JI;LK}
  = \left(
  \begin{array}{l}
    \left[ (\s{q})_{\beta\alpha} (\s{q})_{\delta\gamma}
      + (\s{q} \gamma^5)_{\beta\alpha}(\s{q}\gamma^5)_{\delta\gamma} \right]
    \delta^{JI}\delta^{LK}\\
    \left[ (\s{q})_{\beta\alpha} (\s{q})_{\delta\gamma}
      - (\s{q} \gamma^5)_{\beta\alpha}(\s{q}\gamma^5)_{\delta\gamma} \right]
    \delta^{JI}\delta^{LK}\\
    \left[ (\s{q})_{\beta\gamma} (\s{q})_{\delta\alpha}
      - (\s{q} \gamma^5)_{\beta\gamma}(\s{q}\gamma^5)_{\delta\alpha} \right]
    \delta^{JK}\delta^{LI}
  \end{array}
  \right) \,.
\end{equation} 
The corresponding matrices $F$ read
\begin{equation}
F^{(\gamma^\mu)}=\left(
\begin{array}{ccc}
128N(N+1)   &  0        &    0    \\
0           &  128N^2   &   128N  \\
0           &  128N     &   128N^2 
\end{array}
\right)
\end{equation}
and
\begin{equation}
  F^{\s{q}}=q^2\left(
  \begin{array}{ccc}
    32N(N+1) &  0          &    0      \\
    0           &  32N^2   &   32N  \\
    0           &  32N     &   32N^2  
  \end{array}
  \right)\;,
\end{equation}
 where $N=3$ is the number of colors.

The final result for the amplitude is, of course, independent of the choice of intermediate scheme defined by $P$, but comparing the results obtained with different projection operators gives us an estimate of the systematic uncertainty due to the truncation of perturbation theory in relating the RI-SMOM schemes to the \MSbar schemes. 

The renormalized operators are related to the bare ones by a matrix relation of the form
\begin{equation}
   Q_i^{\!\mathrm{\,R}} (\mu)  = Z_{ij}(\mu a) \,Q_j^\mathrm{bare} (a).
  \label{eq:renormalization}
\end{equation}
In order to extract the renormalization constants we follow the standard procedure~\cite{Martinelli:1994ty,Donini:1999sf} 
and compute numerically the amputated Green's functions of the bare operators in Eq.\,(\ref{eq:4qop})
with particular choices of external momenta (as discussed below) on Landau gauge-fixed configurations. We next solve Eq.~(\ref{eq:renorm_cond}) which we rewrite in the form
\begin{equation}
\label{eq:renorm_cond2}
  \frac{Z_{ij}(\mu a)}{Z_q^2(\mu a)} P_{k}\left\{(\Lambda_j^{\textrm{bare}}(a)\right\}_{\mu^2=p^2} = F_{ik} \;,
\end{equation}
where $\sqrt{Z_q}$ is the quark field renormalization constant and $\mu$ is the renormalization scale, which we ultimately choose to be 3\,GeV.

The choice of $Z_q$ is also not unique, and we use the following two cases:
	\begin{equation}\label{eq:zq}
		\frac{Z_q^{(\s{q})}}{Z_V} = \frac{q^\mu}{12q^2}\mathrm{Tr} \Lambda_V^\mu\s{q},\quad\mathrm{and}\quad
		\frac{Z_q^{(\gamma_\mu)}}{Z_V} = \frac{1}{48}\mathrm{Tr} \Lambda_V^\mu\gamma^\mu,
	\end{equation}
where $\Lambda_V^\mu$ is the three-point amputated Green's function of the local vector current and $Z_V$ is the renormalization constant of the local vector current.
In practice, we multiply each side of Eq.\,(\ref{eq:renorm_cond2}) by the square of the corresponding side of Eq.\,(\ref{eq:zq}). This eliminates $Z_q$ and after this multiplication the left-hand side of Eq.\,(\ref{eq:renorm_cond2}) contains the ratio of renormalization factors
$Z_{ij}/Z_V^2$.
$Z_V$ is then calculated by imposing the Ward identity $Z_V\langle P\,|\,V^4\,|\,P\rangle=2m_P$,  where $V^\mu$ is the local vector current and $|P\rangle$ is the state of a pseudoscalar meson $P$ at rest with mass $m_P$; this is explained in detail in~\cite{Blum:2014tka}.

The choice of projection operator for the four-quark operator and $Z_q$ defines a renormalization scheme, which we will label $(a,b)$ with $a,b \in {\gamma^\mu, \s{q}}$ for the choice of $P^{(a)}$ and $Z_q^{(b)}$.
In particular, we consider the ($\gamma^\mu$,$\gamma^\mu$) and ($\s{q}$,$\s{q}$) schemes, having found in earlier studies that the perturbative conversion to the \MSbar scheme is more precise in these schemes. This is based on the observation that the nonperturbative running is generally closer to the perturbative one for these schemes for the four-quark operators in Eq.\,(\ref{eq:4qop})~\cite{Blum:2012uk,Arthur:2011cn}. As explained below, we follow our previous practice and choose the  ($\s{q}$,$\s{q}$) scheme for our central value and the ($\gamma^\mu$,$\gamma^\mu$) scheme to estimate the error due to the perturbative conversion to the \MSbar scheme.


	\begin{figure}
		\begin{center}
		\includegraphics[scale=1.0]{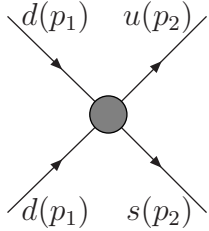}
		\caption{Momentum flow defining a renormalization condition of a four-quark operator in RI-SMOM scheme. The momenta are chosen so that $p_1^2=p_2^2=(p_1-p_2)^2\equiv\mu^2$.}
		\label{fig:rismom}
		\end{center}
	\end{figure}

	Chiral symmetry suppresses mixing of operators in different irreducible representations of the chiral symmetry group, so that if the symmetry is exact, $Z_{ij}$ is a block diagonal matrix with a $1\times 1$ block corresponding to the renormalization of the $(27,1)$ operator and $2 \times 2$ block corresponding to the mixing of $(8,8)$ and $(8,8)_\mathrm{mx}$ operators. In a massless renormalization scheme with a chiral discretization such as the domain-wall action, we expect a mixing pattern very similar to this, but with a small $O((am_\mathrm{res})^2)$ mixing between the blocks.

The mixing of the operator $Q_{(27,1)}$ with either of $Q_{(8,8)}$ or $Q_{(8,8)_\textrm{mx}}$
due to explicit chiral symmetry breaking
induced by finite $L_s$ is proportional to $(am_{\rm res})^2$
(which is $\simle 3.6 \times 10^{-7}$ in this work).
Such mixing can result from two mechanisms~\cite{Aoki:2005ga,Christ:2005xh}.  
First, both quarks in a left-handed $\bar{q}$-$q$ pair in
$Q_{(27,1)}$ can propagate in the fifth dimension from the
left-hand to the right-hand wall, exploiting numerous
but exponentially damped modes which even in perturbation
theory link the left- and right-hand walls.  
This will change
the $(27,1)$ operator into one transforming as the $(8,8)$ representation, but requires the
propagation of two quarks from the left-hand to the
right-hand wall. This incurs a penalty of $(am_{\rm res})^2$
since one power of the residual mass results from the
fifth-dimensional mixing of the left- and right-handed
components of a single quark.  

The second mechanism is nonperturbative and more subtle.
For this case the propagation results from the left-right
tunneling that can be caused by an eigenvector of the
five-dimensional transfer matrix with a near-unit eigenvalue.  
Such eigenvectors permit $O(1)$ left-right mixing but are
rare and therefore give a small contribution to $m_\textrm{res}$.
Under some circumstances such modes can simultaneously
allow a number of quark flavors to flip chirality. However,
to change a (27,1) representation into an (8,8) one, both a quark
and an antiquark must flip chirality which requires two
distinct transfer matrix eigenvectors and is therefore also
doubly suppressed by a  factor $(am_{\rm res})^2$.  Such doubled
suppression will not occur for the mixing between the operator
$Q_{(27,1)}$ and, for example, an operator in the
$(\bar{6},6)$ representation.  Here a single transfer matrix
eigenvector with near-unit eigenvalue can result in
a $O(am_{\rm res})$ mixing between $Q_{(27,1)}$ and
$(\overline{s}(1+\gamma^5)d)\,(\overline{u}(1+\gamma^5)d)$
by allowing both a $u$- and a $d$-quark (localized near this
eigenvector) to flip chirality.  
This kind of mixing has been studied for example in ~\cite{Boyle:2011kn} and it was found to be largely suppressed by our choice of kinematics, as explained below. 

In order to suppress physical infrared chiral-symmetry breaking effects we choose to impose the renormalization conditions with the kinematics indicated in Fig.\,\ref{fig:rismom} with $p_1^2=p_2^2=(p_1-p_2)^2\equiv\mu^2$.
We compute the Green's functions for several momenta and interpolate to $\mu = 3$\,GeV using a quadratic Ansatz. Using partially twisted boundary conditions, we have a good resolution around the targeted momentum.
The momenta in such RI-SMOM schemes are chosen so that there are no ``exceptional" channels, i.e. no channels in which the square of the momenta is small~\cite{Sturm:2009kb}. (This is in contrast with the original RI-MOM scheme~\cite{Martinelli:1994ty,Donini:1999sf}  in which $p_1=p_2$.) We have already checked that with domain-wall fermions and this choice of kinematics the chirally forbidden matrix elements are numerically negligible~\cite{Blum:2012uk}. In the present computation, we use the $48^3$ and $64^3$
ensembles which have physical light and strange sea-quark
masses.  However, the light-quark mass is used in all
of the valence-quark propagators in the five-point
Green's functions, including those for both light and
strange quarks.  We do not extrapolate either the sea-
or valence-quark masses to zero and, strictly speaking,
do not work in the chiral limit.  In practice the light-quark masses are sufficiently small that their effects
are negligible as is the nonzero mass of the strange
sea quark. Comparing our results with those of our previous work (with Shamir domain-wall fermions) where a chiral extrapolation was performed we find agreement at the per-mille level or better.
                
        We find that all the chirally forbidden renormalization factors are smaller than $10^{-5}$, 
        so we set the corresponding matrix elements of $P_i\{\Lambda_j\}$ to zero and finally obtain the renormalization matrices:
        \begin{eqnarray}
          \label{numbersZ1}
		Z^{(\gamma^\mu,\gamma^\mu)}_{\beta=2.13}(\mu=3\, \mathrm{GeV}) &=& \left( \begin{array}{ccc}
				0.4617(3) & 0 & 0 \\
				0 & 0.5302(4) & -0.07018(6) \\
				0 & -0.0386(1) & 0.4451(5) 
			\end{array}
		\right)\\[0.2in]
          \label{numbersZ2}
		Z^{(\s{q},\s{q})}_{\beta=2.13}(\mu=3\,\mathrm{GeV}) &=& \left( \begin{array}{ccc}
				0.4822(3) & 0 & 0 \\
				0 & 0.5305(4) & -0.07135(7)\\
				0 & -0.0637(1) & 0.5052(6) 
			\end{array}
		\right)\end{eqnarray}
  for the $48^3$ ensembles and   
\begin{eqnarray}
                \label{numbersZ3}
		Z^{(\gamma^\mu,\gamma^\mu)}_{\beta=2.25}(\mu=3\, \mathrm{GeV})&=& \left( \begin{array}{ccc}
				0.5194(2) & 0 & 0 \\
				0 & 0.5774(2) & -0.0751(1)\\
				0 & -0.02797(7) & 0.4431(6) 
			\end{array}
		\right)\\[0.2in]
                \label{numbersZ4}
		Z^{(\s{q},\s{q})}_{\beta=2.25}(\mu=3\,\mathrm{GeV}) &=& \left( \begin{array}{ccc}
				0.5399(2) & 0 & 0 \\
				0 & 0.5782(2) & -0.0761(1) \\
				0 & -0.05230(4) & 0.4990(5)
			\end{array}
		\right)
	\end{eqnarray}
for the $64^3$ ensembles.
With momentum sources~\cite{Gockeler:1998ye}, only a few configurations are needed to obtain an excellent statistical precision. The number of Landau gauge-fixed configurations used to obtain these results varies between 5 and 15. The statistical errors were estimated with 200 bootstrap samples.
The matrices in Eqs.\,(\ref{numbersZ1})\,--\,(\ref{numbersZ4}) are the ones used in Eq.\,(\ref{eq:renormalization}) to obtain the operators renormalized in the RI-SMOM schemes at the scale $\mu=3\,\mathrm{GeV}$ from the corresponding lattice bare operators.

The procedure described above enables us to calculate the matrix elements of the operators in Eq.\,(\ref{eq:4qop}) in the (continuum) RI-SMOM schemes with a very small systematic uncertainty due to the renormalization. The Wilson coefficients however, are computed in the \MSbar scheme and so we have to match the RI-SMOM schemes to the \MSbar one. We repeat that this matching is perturbative and at present is only known to one-loop order~\cite{Lehner:2011fz}; this limitation amplifies the uncertainty due to the renormalization. This uncertainty could be reduced by extending the perturbative calculations to higher orders. Future lattice calculations could also help here by using step scaling to run the renormalization constants obtained in the RI-SMOM schemes nonperturbatively to larger momentum scales. 
The perturbative matching to the \MSbar scheme  can then be performed at these larger scales where the coupling constant is smaller, leading to smaller uncertainties.
We now estimate the current uncertainty due to the matching.

	\begin{table}[t]
		\begin{center}
		\begin{tabular}{|c|c|c|}
		\hline
		& $48^3$ ensembles & $64^3$ ensembles		\\
		\hline
		Re($A_2$) $(\gamma^\mu, \gamma^\mu)$ & $1.346(11)_{\rm{stat}}(1)_{\rm{NPR}} \times 10^{-8}$\,GeV & $1.4029(93)_{\rm{stat}}(11)_{\rm{NPR}} \times 10^{-8}$\,GeV\\
		Im($A_2$) $(\gamma^\mu, \gamma^\mu)$ & $-5.739(46)_{\rm{stat}}(8)_{\rm{NPR}} \times 10^{-13}$\,GeV& $-6.143(73)_{\rm{stat}}(9)_{\rm{NPR}} \times 10^{-13}$\,GeV\\
		Re($A_2$) $(\s{q},\s{q})$ & $1.386(12)_{\rm{stat}}(1)_{\rm{NPR}} \times 10^{-8}$\,GeV & $1.4386(95)_{\rm{stat}}(11)_{\rm{NPR}} \times 10^{-8}$\,GeV \\
		Im($A_2$) $(\s{q},\s{q})$ & $-6.174(49)_{\rm{stat}}(9)_{\rm{NPR}} \times 10^{-13}$\,GeV&$-6.548(78)_{\rm{stat}}(10)_{\rm{NPR}} \times 10^{-13}$\,GeV \\
		\hline
		\end{tabular}
		\caption{The amplitude $A_2$ calculated using two different intermediate RI-SMOM schemes. The two errors, labeled ``stat" and ``NPR"  are the statistical uncertainties in the evaluation of the bare matrix elements and $Z_{ij}$ respectively.
Discrepancies in the results in the two schemes are attributed to the truncation in the matching to the 
$\overline{\mathrm{MS}}$ scheme.}
		\label{tab:a2int}
		\end{center}
	\end{table}

To estimate the uncertainty due to the truncation of the perturbative matching factors,  we note that the matrix elements in the \MSbar scheme should be independent of the choice of intermediate RI-SMOM scheme. Differences in the results are observed (see Table~\ref{tab:a2int}) and attributed to the truncation. Following the procedure in~\cite{Blum:2011ng,Blum:2012uk} we take the result obtained using the $(\s{q},\s{q})$ intermediate scheme as our central value and the difference of the results obtained using the two schemes as an estimate of the systematic error. This uncertainty is marked as ``NPR (perturbative)" in the error budgets presented in Tables~\ref{tab:errre} and \ref{tab:errim} in Sec.\,\ref{sec:errors}. The uncertainties marked as ``NPR (nonperturbative)" are the \rm{stat}istical errors in the evaluation of $Z_{ij}$.

\section{Finite-volume effects}\label{sec:FV}
	
The presence of two pions in the final \rm{stat}e in $K\to\pi\pi$ decays leads to finite-volume corrections which decrease as inverse powers of the volume, in addition to the exponential correction present in simpler quantities such as decay constants and form factors. The power corrections result in a multiplicative correction to the matrix element \cite{Lellouch:2000pv}:
	\begin{equation}\label{eq:LLrelation}
		\langle \pi\pi \mid H_W \mid K \rangle_\infty = F \langle \pi\pi \mid H_W \mid K \rangle_{FV}\:.
	\end{equation}
	The subscripts $\infty$ and $FV$ correspond to infinite and finite volume respectively, and the factor $F$ is given by the Lellouch-L\"{u}scher formula~\cite{Lellouch:2000pv}:
	\begin{equation}\label{eq:Fsq}
		F^2 = 8\pi q \left(\frac{\partial \phi}{\partial q} +  \frac{\partial \delta}{\partial q} \right)\frac{m_KE_{\pi\pi}^2}{p^3}\:,
	\end{equation}
	where $p$ is the magnitude of the momentum of a pion in the center-of-mass frame given by $p=\sqrt{\frac{E_{\pi\pi}^2}{4}-m_\pi^2}$ and $q$ is defined as 
	$q=pL/2\pi$. 
	Since the $\pi^+$ mesons satisfy antiperiodic boundary conditions in all three spatial directions, the function $\phi$ in this case is defined by the condition:
	\begin{equation}
		\tan \phi = -\frac{q \pi^{3/2} }{Z_{00}(1;q)},\quad Z_{00}(1;q) = \frac{1}{\sqrt{4\pi}} \sum_{n \in \mathbb Z^3} \frac{1}{(n+(\frac12,\frac12,\frac12))^2-q^2}\:.
	\end{equation}
	$\delta$ is the two-pion s-wave phase shift, which can be calculated using the L\"{u}scher quantization condition, $\delta(q) + \phi(q) = n\pi$, but the calculation of the derivative in Eq.\,(\ref{eq:Fsq}) requires an approximation.
	
	\begin{table}[t]
	\begin{center}
	\begin{tabular}{c|c|c|c|c|c}
	&$E_{\pi\pi}$&  q & $\delta$ (radians) & $\frac{\partial \delta}{\partial q}$ & $\frac{\partial \phi}{\partial q}$ \\
	\hline
	$48^3$ & 0.2873(13) & 0.9087(61) & $-0.158(22)$ & $-0.174(24)$ & 3.7147(20)\\
	$64^3$ & 0.21512(68) & 0.9157(43) & $-0.184(16)$ & $-0.201(17)$ & 3.7171(15)
	\end{tabular}
	\end{center}
	\caption{Contributions to the Lellouch-L\"{u}scher factor on the $48^3$ and $64^3$ ensembles. The rate of change of the phase shift was calculated by using a linear approximation in momentum as explained in the text.}
	\label{tab:llfactor}
	\end{table}

\begin{figure}
	\begin{center}
		\includegraphics[scale=0.6]{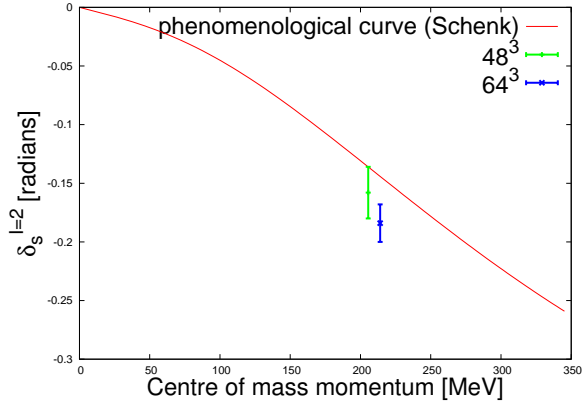}
		\caption{Comparison of $I=2$ two-pion s-wave phase shifts calculated using L\"{u}scher's formula with the phenomenological curve from Ref.\,\cite{Schenk:1991xe}. The computed results are consistent with the phenomenological curve.
		\label{fig:schenk}}
	\end{center}
\end{figure}

	The results presented in Table~\ref{tab:llfactor} were obtained using the approximation that $\delta$ is a linear function of the momentum between $0$ and $p$. Since the second term in the parentheses on the right-hand side of Eq.\,(\ref{eq:Fsq}) is much smaller than the first and given the remaining systematic uncertainties discussed in Sec.\,\ref{sec:errors}, this procedure gives an adequate approximation. In order to estimate the error due to this approximation we also evaluate the derivative $\frac{\partial \delta}{\partial p}$
using the phenomenological curve of Ref.\,\cite{Schenk:1991xe} illustrated in Fig.\,\ref{fig:schenk}; we take the difference of the two procedures as an estimate of the corresponding uncertainty. 
For our central value we use the linear approximation for the derivative of the phase shift so that it is independent of phenomenological estimates.

At the pion momentum which corresponds to the decay of a physical kaon to two pions ($p=207 \, \mathrm{MeV}$) the value of the derivative of the phase shift with respect to the momentum obtained from the phenomenological curve is $9.53\times 10^{-4}\,\mathrm{MeV}^{-1}$.
Converting this to $\frac{\partial \delta}{\partial q}$ gives $-0.216$ for the $48^3$ and $-0.221$ for the $64^3$ ensembles.
While this makes a significant difference to the derivative of the phase shift, it represents a relatively small uncertainty in the Lellouch-L\"uscher factor $F \propto \frac{\partial \delta}{\partial q} + \frac{\partial \phi}{\partial q}$.
This sum is dominated by the $\frac{\partial \phi}{\partial q}$ term and thus the difference in the Lellouch-L\"{u}scher factor between both approaches to calculating $\frac{\partial \delta}{\partial q}$ amounts to 1.1\% and 0.6\% on the $48^3$ and $64^3$ ensembles respectively.

When quoting our central value we include the Lellouch-L\"uscher factor evaluated as described in the preceding paragraph. In order to estimate the size of the remaining exponential finite-volume effects we use chiral perturbation theory and include the corresponding effects in our systematic uncertainty. Since we are only calculating an estimate, we do not use partially twisted chiral perturbation theory, but take both the sea and valence $d$-quarks to satisfy antiperiodic boundary conditions. 

In SU(3)$_\mathrm{L}\times$SU(3)$_\mathrm{R}$ chiral perturbation theory, the leading order (LO) and leading logarithmic next-to-leading order (log) contributions to the (27,1) and (8,8) matrix elements are given by \cite{Aubin:2008vh,Laiho:2002jq}
\begin{align}
\mathcal{M}^{27}_\text{LO}=&\langle\pi^+\pi^-|\mathcal{O}^{(27,1),3/2}|K^0\rangle_\text{LO} = -\frac{4i\alpha_{27}}{f_Kf_\pi^2}(m_K^2-m_\pi^2)\,, \label{O27_LO} \\
\mathcal{M}^{27}_\text{log}=&\langle\pi^+\pi^-|\mathcal{O}^{(27,1),3/2}|K^0\rangle_\text{log} \nonumber \\
= & -\frac{4i\alpha_{27}}{f_Kf_\pi^2}\frac{1}{f^2}\left[-\frac{1}{12}m_K^4\left(1-\frac{m_K^2}{m_\pi^2}\right)\beta(m_\pi^2,m_K^2,m_\eta^2)+m_K^2\left(\frac{5}{4}\frac{m_K^4}{m_\pi^2}\right.\right. \nonumber \\
&\left.-\frac{13}{4}m_K^2+2m_\pi^2\right)\beta(m_\pi^2,m_K^2,m_\pi^2)+(m_K^4-3m_\pi^2m_K^2+2m_\pi^4) \nonumber \\
&\times\beta(m_K^2,m_\pi^2,m_\pi^2)+\left(-\frac{1}{4}\frac{m_K^4}{m_\pi^2}-\frac{1}{12}m_K^2+\frac{1}{3}m_\pi^2\right)\ell(m_\eta^2)+\left(\frac{-m_K^4}{m_\pi^2} \right. \nonumber \\
& \left. \left. -4m_K^2+4m_\pi^2\right)\ell(m_K^2)+\left(\frac{5}{4}\frac{m_K^4}{m_\pi^2}-\frac{45}{4}m_K^2+11m_\pi^2\right)\ell(m_\pi^2)\right] \,,\label{O27_log} \\
\mathcal{M}^{88}_\text{LO}=&\langle\pi^+\pi^-|\mathcal{O}^{(8,8),3/2}|K^0\rangle_\text{LO} = -\frac{4i\alpha_{88}}{f_Kf_\pi^2}\,, \label{O88_LO)} \\
\mathcal{M}^{88}_\text{log}=&\langle\pi^+\pi^-|\mathcal{O}^{(8,8),3/2}|K^0\rangle_\text{log} \nonumber \\
= & -\frac{4i\alpha_{88}}{f_Kf_\pi^2}\frac{1}{f^2}\left[\left(\frac{5}{4}\frac{m_K^4}{m_\pi^2}-2m_K^2\right)\beta(m_\pi^2,m_K^2,m_\pi^2)+(m_K^2-2m_\pi^2) \right. \nonumber \\
&\left. \times \beta(m_K^2,m_\pi^2,m_\pi^2)+\frac{1}{4}\frac{m_K^4}{m_\pi^2}\beta(m_\pi^2,m_K^2,m_\eta^2)-\left(4+\frac{1}{2}\frac{m_K^2}{m_\pi^2}\right)\ell(m_K^2) \right. \nonumber \\
&\left. +\left(\frac{5}{4}\frac{m_K^2}{m_\pi^2}-8\right)\ell(m_\pi^2)-\frac{3}{4}\frac{m_K^2}{m_\pi^2}\ell(m_\eta^2)\right]. \label{O88_log}
\end{align}
At this order $m_\eta$ is given by the Gell-Mann-Okubo relation: $3 m_\eta^2 = 4m_K^2-m_\pi^2$.

\begin{figure}
	\begin{subfigure}[c]{0.3\textwidth}
	\includegraphics[width=\textwidth]{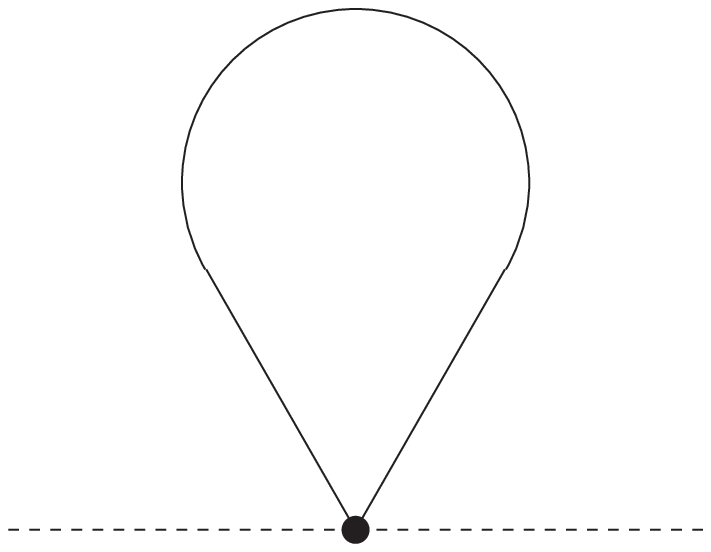}
	\end{subfigure}
	\qquad
	\begin{subfigure}[c]{0.45\textwidth}
	\includegraphics[width=\textwidth]{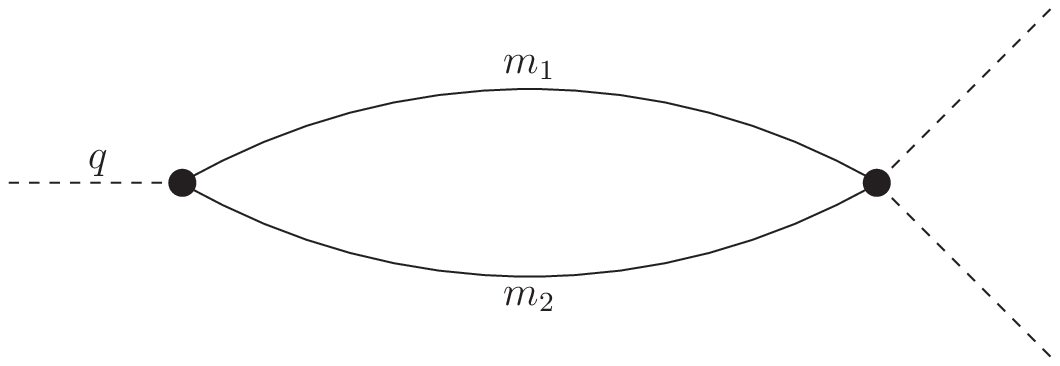}
	\end{subfigure}
	\caption{Sample loop diagrams which contribute to finite-volume corrections of (27,1) and (8,8) $K \rightarrow \pi\pi$ matrix elements in chiral perturbation theory.}
	\label{fig:chpt}
\end{figure}

The functions $\ell(m^2)$ and $\beta(q^2,m_1^2,m_2^2)$ correspond to diagrams with one and two pseudo Goldstone boson propagators respectively as illustrated in Fig.\,\ref{fig:chpt} and they are the only sources of finite-volume corrections.
They are given by (in Minkowski spacetime)
\begin{align}
	\ell(m^2) &\equiv \SumInt \int \frac{dk^0}{2\pi} \frac{i}{k^2 - m^2 + i\epsilon}=\SumInt\frac{1}{\sqrt{\vec{k}^2+m^2}}\,,\label{eq:ldef}\\ 
\beta(q,m_1,m_2) &\equiv \SumInt \int \frac{dk^0}{2\pi} \frac{i}{\left(k^2-m_1^2\right)\left((q+k)^2-m_2^2\right)}=\SumInt\frac{\omega_1+\omega_2}{2\omega_1\omega_2(q_0^2-(\omega_1+\omega_2)^2)}\,,\label{eq:betadef}
\end{align}
where the symbol $\SumInt$ denotes the summation over $\vec{k}$ in finite volume or the integration in infinite volume. $\omega_1=\sqrt{\vec{k}^2+m_1^2}$ and $\omega_2=\sqrt{(\vec{q}-\vec{k})^2+m_2^2}$.
The difference between the sum and the integral can be calculated using the Poisson summation formula:
\begin{equation}\label{eq:poisson}
	\frac{1}{L^3} \sum_{\vec k} f(\vec k) = \int \frac{d^3 \vec k}{(2\pi)^3} f(\vec k) + \sum_{\vec n \neq \vec 0} \int \frac{d^3 \vec k}{(2\pi)^3} f(\vec k) e^{iL \vec k \cdot \vec n},
\end{equation}where the summation on the left-hand side is over all $\vec{k}=\frac{2\pi}{L}\vec{n}$, where $\vec{n}$ is a vector of integers.
If $f$ is a function that has no singularities on the real axis, then the second term on the right-hand side gives the exponential 
finite-volume corrections which we are trying to evaluate. 

\subsection{Corrections to $\boldsymbol{\ell(m^2)}$}
With periodic boundary conditions, applying the Poisson summation formula (\ref{eq:poisson}) to $\ell$, writing $\vec k$ in spherical polar coordinates and integrating over the angles, we obtain for the difference between the finite- and infinite-volume values of $\ell(m^2)$\,\cite{Aubin:2003mg}
\begin{equation}\label{eq:l_correction}
\Delta \ell(m,L)\equiv\frac{m^2}{16\pi^2}\delta_1(mL) \equiv \frac{m}{4\pi^2 L}\sum\limits_{\vec{n}\ne 0} \frac{K_1(|\vec{n}|mL)}{|\vec{n}|}\,,
\end{equation}
where 
$K_1$ is a modified Bessel function of the second kind, $\vec{n}$ is an vector of integers and the sum is over all $\vec{n}\neq(0,0,0) \in \mathbb Z^3$.   

Since our choice of boundary conditions breaks the isospin symmetry Eq.\,(\ref{eq:l_correction}) does not give the correct finite-volume corrections for all the instances of $\ell$ which appear in Eqs.\,(\ref{O27_log}) and (\ref{O88_log}). Specifically, $\pi^0$, $K^+$ and $\eta$ satisfy periodic boundary conditions (so that the corresponding finite-volume corrections are indeed given by Eq.\,(\ref{eq:l_correction})) whereas $\pi^\pm$ and $K^0$ satisfy antiperiodic boundary conditions for which the finite-volume corrections to $\ell$ are different.
In the antiperiodic case, we replace $f(\vec k)$ in Eq.\,(\ref{eq:poisson}) by $f(\vec k + \vec q)$, where $\vec q = (\frac{\pi}{L})(1,1,1)$.
Shifting the integration variable from $\vec k$ to $\vec k + \vec q$, we find that $\delta_1(mL)$ in Eq.\,(\ref{eq:l_correction}) is now replaced by
\begin{equation}
\delta_1^A(mL)=\frac{4}{mL}\sum\limits_{\vec{n}\ne \vec 0} (-1)^{n_x+n_y+n_z}\frac{K_1(|\vec{n}|mL)}{|\vec{n}|}\,,
\end{equation}
where the index $A$ denotes that the correction is evaluated for a volume with antiperiodic boundary conditions in all spatial directions.
The difference from the periodic case is the additional factor $(-1)^{n_x+n_y+n_z}$ in the summands.
The known formulas in Eqs.\,(\ref{O27_log}) and (\ref{O88_log}) do not differentiate between different isospin components, and therefore do not specify which linear combination of periodic and antiperiodic corrections should be used. Since we are only using these formulas for an approximate estimate of the size of the error, we choose to be conservative and to include the larger corrections which are those obtained with the periodic boundary conditions given in Eqs.\,(\ref{eq:l_correction}). The numerical results are presented in Table~\ref{tab:fvexp} and as expected the leading contributions come from the loops with a pion propagator.

\subsection{Corrections to $\boldsymbol{\beta(m_\pi,m_K,m_\pi)}$ and $\boldsymbol{\beta(m_\pi,m_K,m_\eta)}$}\label{subsec:beta1}
We now consider the contributions from loops with two meson propagators and which are proportional to the function $\beta$. We start by discussing the corrections to $\beta(m_\pi,m_K,m_\pi)$ and $\beta(m_\pi,m_K,m_\eta)$, for which in Minkowski space the external energy is below the corresponding two-particle cut; e.g. in $\beta(m_\pi,m_K,m_\pi)$ the external energy in the center-of-mass frame is $m_\pi$ which is clearly smaller than $m_K+m_\pi$. In such situations the finite-volume corrections are exponentially small. We postpone the discussion of the contribution which does contain the two-particle cut, that proportional to $\beta(m_K,m_\pi,m_\pi)$, until the following subsection.

The corrections to $\beta(m_\pi,m_K,m_\pi)$ and $\beta(m_\pi,m_K,m_\eta)$ are proportional to 
\begin{equation}\label{eq:Deltabeta}
	\Delta \beta(q,m_1,m_2) = \sum_{\vec n \neq 0} \int \frac{d^3 \vec k}{(2\pi^3)} \frac{e^{i\vec k \cdot \vec n}(\omega_1+\omega_2)}{2\omega_1\omega_2(q^2-(\omega_1+\omega_2)^2)}
\end{equation} 
with
\begin{equation}
\omega_1^2 = \left|\vec k \right|^2 + m_1^2\qquad\mathrm{and}\qquad
\omega_2^2 = \left|\vec q+ \vec k \right|^2 + m_2^2\,.
\end{equation}
Because of the angular dependence inside the integrals, we evaluate the integrals numerically.
With the boundary conditions which we are using the corrections with a $K^+$ and $\pi^-$ are equal and opposite to those with the neutral mesons. In the estimate of the uncertainty we conservatively do not exploit the cancellation but take the absolute value in each case. 

We note that care must be taken 
when using Eqs.\,(71) and (73) for the finite-volume corrections to $\beta$ in Sec.\,VIII of~\cite{Aubin:2008vh}.  In Eq.\,(\ref{eq:Deltabeta}) above, the two terms in the factor in the denominator of the integrand $q^2-(\omega_1+\omega_2)^2$ come with opposite signs. How this arises in finite-volume Euclidean correlation functions is explained in the Appendix following~\cite{Lin:2002nq}. The corresponding terms in the denominator of Eq.\,(73) in~\cite{Aubin:2008vh} appear (incorrectly) with the same sign. 

\subsection{$\boldsymbol{\beta(m_K, m_\pi, m_\pi)}$}\label{subsec:beta2}
Kinematically this case is simpler than the two $\beta$ integrals which were evaluated in Sec.\ref{subsec:beta1} since the external particle ($K$) is now at rest which eliminates the angular dependence from the integral.
Furthermore, both internal $\pi^+$ propagators satisfy antiperiodic boundary conditions. 
In this case however, the integral for $\beta$ has a pole at $\omega_\pi = m_K/2$, so the Poisson summation formula will give both the exponential and powerlike corrections. The power corrections are included as the Lellouch-L\"uscher factor $F$ in Eq.\,(\ref{eq:LLrelation}) and we do not include these in the estimate of the finite-volume uncertainty. The evaluation of the remaining exponential corrections following the approach of~\cite{Kim:2005gf} is explained in the Appendix.

\subsection{Combining the finite-volume corrections}
To one-loop order we write the systematic error associated with the finite-volume corrections in terms of the ratios $\Delta \mathcal M_\text{log}/\mathcal{M}_\text{LO}$.
These are given by:
\begin{eqnarray}
\frac{\Delta \mathcal{M}^{27}_\text{log}}{\mathcal{M}^{27}_\text{LO}} & = & \frac{1}{f^2(m_K^2-m_\pi^2)}\left[-\frac{1}{12}m_K^4\left(1-\frac{m_K^2}{m_\pi^2}\right)\Delta\beta(m_\pi^2,m_K^2,m_\eta^2)\right. \nonumber \\
&&\hspace{-0.5in}+m_K^2\left(\frac{5}{4}\frac{m_K^4}{m_\pi^2}-\frac{13}{4}m_K^2+2m_\pi^2\right)\Delta\beta(m_\pi^2,m_K^2,m_\pi^2)+\nonumber \\
&&\hspace{-0.8in}(m_K^4 -3m_\pi^2m_K^2+2m_\pi^4)\Delta\beta(m_K^2,m_\pi^2,m_\pi^2)+\left(-\frac{1}{4}\frac{m_K^4}{m_\pi^2}-\frac{1}{12}m_K^2  +\frac{1}{3}m_\pi^2\right)\Delta\ell(m_\eta^2)
\nonumber \\
&& \hspace{-0.8in}\left.+\left(\frac{-m_K^4}{m_\pi^2}-4m_K^2+4m_\pi^2\right)\Delta\ell(m_K^2)
+\left(\frac{5}{4}\frac{m_K^4}{m_\pi^2}-\frac{45}{4}m_K^2+11m_\pi^2\right)\Delta\ell(m_\pi^2)\right] \label{O27_correction}
\end{eqnarray}
and
\begin{eqnarray}
\frac{\Delta \mathcal{M}^{88}_\text{log}}{\mathcal{M}^{88}_\text{LO}}& = & \frac{1}{f^2}\left[\left(\frac{5}{4}\frac{m_K^4}{m_\pi^2}-2m_K^2\right)\Delta\beta(m_\pi^2,m_K^2,m_\pi^2)+(m_K^2-2m_\pi^2) \Delta\beta(m_K^2,m_\pi^2,m_\pi^2)\right. \nonumber \\
&&\hspace{-0.3in}+\frac{1}{4}\frac{m_K^4}{m_\pi^2}\Delta\beta(m_\pi^2,m_K^2,m_\eta^2)-\left(4+\frac{1}{2}\frac{m_K^2}{m_\pi^2}\right)\Delta\ell(m_K^2) \nonumber \\
&&\hspace{0.5in}\left.
+\left(\frac{5}{4}\frac{m_K^2}{m_\pi^2}-8\right)\Delta\ell(m_\pi^2)-\frac{3}{4}\frac{m_K^2}{m_\pi^2}\Delta\ell(m_\eta^2)\right]\,. \label{O88_correction} 
\end{eqnarray}
The numerical values of these ratios for the $48^3$ and $64^3$ ensembles are shown in Table~\ref{tab:fvexp}.
	\begin{table}
	\begin{center}
	\begin{tabular}{|c|c|c|}
		\hline
		 Quantity &  $48^3$~lattice & $64^3$~lattice \\
		\hline
		$L$ & 5.48\,fm & 5.36\,fm\\
		$\Delta l(m_\pi^2)$ & $14.32\, \MeV^2$ & $16.39\, \MeV^2$ \\
		$\Delta l(m_K^2)$ & $~(9.05\times 10^{-4})\, \MeV^2$~~& ~$(1.03\times 10^{-3})\, \MeV^2$~~\\
		$\Delta l(m_\eta^2)$ &~$(1.32\times 10^{-4})\, \MeV^2$~~& ~$(1.52\times 10^{-4})\, \MeV^2$~~\\
		$\Delta \beta(m_\pi,m_K,m_\eta)$  & $3.0 \times 10^{-7}$ & $3.0 \times 10^{-7}$\\
		$\Delta \beta(m_\pi,m_K,m_\pi)$ & $5.0\times 10^{-5}$ & $5.2 \times 10^{-5}$\\
		$\Delta \beta(m_K,m_\pi,m_\pi)$ & $6.67 \times 10^{-5}$ & $6.97 \times 10^{-5}$ \\
		\hline
		$\frac{\Delta M_{(27,1)}}{M_{(27,1)}}$ & 0.022 & 0.024\\
		$\frac{\Delta M_{(8,8)}}{M_{(8,8)}}$ & 0.024 & 0.026\\
		\hline
	\end{tabular}
	\end{center}
		\caption{Contributions to our estimate of the exponentially suppressed
finite-volume errors.}
		\label{tab:fvexp}
	\end{table}
	
\section{The Error budget}\label{sec:errors}
In this section we discuss the two remaining systematic errors: those which arise because the meson masses and the two-pion energy are not quite physical and those introduced by the perturbative Wilson coefficients.  Finally all of the systematic errors in our results for the real and imaginary parts of $A_2$ are summarized in Tables \ref{tab:errre} and \ref{tab:errim}, respectively.
	
The volume, boundary conditions and quark masses have been chosen to enable simulations of physical $K\to\pi\pi$ decays. Nevertheless, since the volume and quark masses have to be chosen \emph{a priori}, the output values of the meson masses and two-pion energies will be a little different from the physical values (see Table~\ref{tab:masse}). In order to estimate the corresponding uncertainty we follow the 
procedure described in \cite{matthew,Blum:2012uk} and outlined below. We use measurements on 60 quenched configurations on a $24^3$ lattice with $a^{-1}=1.31\,\mathrm{GeV}$ performed with three values of the light-quark masses, five strange-quark masses and the application of antiperiodic boundary conditions in $n_\mathrm{tw}$\,=\,0, 1, 2 and 3 directions.
	These measurements are used to determine the coefficients in the following phenomenological formulas:
	\begin{align}
		m_{xy}^2 &= B_0(m_x + m_y) + B_1,\\
		E_{\pi\pi}^2(n_{\mathrm{tw}}) &= A_0(n_{\mathrm{tw}}) m_l + A_1(n_{\mathrm{tw}}),\\
		A_2 &= C_0(n_{\mathrm{tw}}) m_s + C_1(n_{\mathrm{tw}})m_l + C_2(n_{\mathrm{tw}}),
	\end{align}
	where $m_l$ and $m_s$ are the masses of the light and strange quarks, $m_{xy}$ is the mass of the meson consisting of $x$ and $y$ valence quarks (which can be either light or strange) and $n_{\mathrm{tw}}$ is the number of directions in which the antiperiodic boundary conditions would have to be imposed on the quenched lattice to get the correct two-pion energy. Note that $n_{\mathrm{tw}}$ does not have to be an integer, and is given instead by $p^2=n_{\mathrm{tw}}\pi^2/L^2$, where $p$ is the center-of-mass momentum of each pion. The full list of coefficients A, B and C obtained from these quenched configurations was presented in \cite{matthew} and is reproduced in Table~\ref{tab:quenchedparams}.

We can use the coefficients in Table~\ref{tab:quenchedparams} to determine $A_2$ on the quenched ensembles for any choice of $\{m_\pi,m_K,E_{\pi\pi}\}$. We exploit this possibility for three sets of parameters: (i)~the physical masses $m_K=E_{\pi\pi}=493.7$\,MeV, $m_\pi=139.6$\,MeV; (ii)~the values from the $48^3$ simulation given in the third row of Table~\ref{tab:masse} and (iii)~the values from the $64^3$ simulation given in the fourth row of Table~\ref{tab:masse}. We denote the corresponding 
three estimates of $A_2$ by $A_2^{\mathrm{q;phys}}$, $A_2^{\mathrm{q};48}$ and $A_2^{\mathrm{q};64}$ respectively, where the superscript $\mathrm{q}$ reminds us that the results were obtained on the quenched ensembles. We use the differences $A_2^{\mathrm{q};48}-A_2^{\mathrm{q;phys}}$ and 
$A_2^{\mathrm{q};64}-A_2^{\mathrm{q;phys}}$	as estimates of the 
systematic error due to unphysical kinematics.
	
	The results are:
\begin{eqnarray}\label{eq:quenched1}
		\mathrm{Re}(A_2^{\mathrm{q;phys}}) &= 2.25\times 10^{-8}\, \mathrm{GeV},\quad
		\mathrm{Im}(A_2^{\mathrm{q;phys}}) &= -1.344 \times 10^{-12}\, \mathrm{GeV},\\
\mathrm{Re}(A_2^{\mathrm{q};48}) &= 2.29\times 10^{-8}\, \mathrm{GeV},\quad
		\mathrm{Im}(A_2^{\mathrm{q};48}) &= -1.341 \times 10^{-12}\, \mathrm{GeV},
	\label{eq:quenched2}\\
\mathrm{Re}(A_2^{\mathrm{q};64}) &= 2.36\times 10^{-8}\, \mathrm{GeV},\quad
		\mathrm{Im}(A_2^{\mathrm{q};64}) &= -1.329 \times 10^{-12}\, \mathrm{GeV}.
		\label{eq:quenched3}
\end{eqnarray}
The differences in Eqs.\,(\ref{eq:quenched1})\,-\,(\ref{eq:quenched3}) translate to an estimated 
$1.8\%$ error on $\mathrm{Re}(A_2)$ and $0.2\%$ error on $\mathrm{Im}(A_2)$ on the $48^3$ ensemble and a $4.5\%$ difference for Re($A_2$) and $1.1\%$ difference for Im($A_2$) on the $64^3$ ensemble. These numbers are obtained from the difference of the simulated results from those at the physical point (normalized by the result at the physical point). These uncertainties are included in Tables~\ref{tab:errre} and \ref{tab:errim} under the label ``unphysical kinematics.''
\begin{table}
	\begin{tabular}{|c|c|c|c|c|}
	\hline
	$n_{\mathrm{tw}}$ & $0$ & $1$ & $2$ & $3$\\
	\hline
	$A_0$ & 17.53(16) & 17.14(73)  & 14.9(2.3) & 24.5(9.5) \\ 
	$A_1$ & 0.0273(12) & 0.1038(60) & 0.202(18) & 0.196(82) \\ 
	$B_0$ & 2.124& 2.124&2.124 &2.124 \\ 
	$B_1$ & 0.00692& 0.00692 & 0.00692 & 0.00692 \\ 
	Re$C_0$(GeV) & $1.016(55) \times 10^{-7}$& $1.43(11)\times 10^{-7}$ & $1.53(25) \times 10^{-7}$ & $1.78(54)\times 10^{-7}$ \\ 
	Re$C_1$(GeV) & $1.697(89) \times 10^{-6}$ & $1.29(18) \times 10^{-6}$ & $1.45(38)\times 10^{-6}$ & $4.22(97)\times 10^{-6}$ \\ 
	Re$C_2$(GeV) & $2.53(51)\times 10^{-9}$ & $1.08(12)\times 10^{-8}$ & $1.68(25)\times 10^{-8}$ & $-2(67)\times 10^{-10}$ \\ 
	Im$C_0$(GeV) & $-1.06(31) \times 10^{-12}$& $-4.6(3.3) \times 10^{-13}$ & $4.4(7.4) \times 10^{-13}$ & $2(11)\times 10^{-13}$ \\ 
	Im$C_1$(GeV) & $5.54(79) \times 10^{-11}$ & $3.39(91) \times 10^{-11}$ & $2.1(1.6) \times 10^{-11}$ & $-1.8(3.2) \times 10^{-11}$ \\ 
	Im$C_2$(GeV) & $-1.689(64) \times 10^{-12}$ & $-1.392(66)\times 10^{-12}$ & $-1.24(12) \times 10^{-12}$ & $-7.5(1.9) \times 10^{-13}$ \\ 
	\hline
	\end{tabular}
	\caption{Parameters used for extrapolations on the $24^3$ quenched ensembles.}
	\label{tab:quenchedparams}
	\end{table}

	\begin{table}[t]
	\begin{tabular}{|c|c|c|c|}
		\hline
		& (27,1) & (8,8) & $(8,8)_{\textrm{mx}}$\\
		\hline
		$z_i^{LO}$ & $0.26696$ & $4.260055 \times 10^{-5}$ & $-1.0063 \times 10^{-5}$ \\
		$y_i^{LO}$ & $-0.0035185$ & $-2.026445 \times 10^{-4}$ & $2.447741 \times 10^{-4}$ \\
		$z_i^{NLO}$ & $0.290342$ & $4.70099 \times 10^{-5}$ & $-5.22390 \times 10^{-5}$ \\
		$y_i^{NLO}$ & $-0.00397252$ & $-8.09555\times 10^{-5}$ & $3.26016\times 10^{-4}$ \\
		\hline
	\end{tabular}
\caption{Wilson coefficients at 3\,GeV in the \MSbar scheme at leading order (LO) and next-to-leading order (NLO).}
	\label{tab:wilson}
	\end{table}

\begin{table}
	\begin{tabular}{|c|c|c|}
	\hline
		& LO & NLO\\
	\hline
	Re($A_2$) $48^3$ & $1.293(11)\times 10^{-8}$ & $1.386(12)\times 10^{-8}$\\
	Im($A_2$) $48^3$ & $-5.551(45)\times 10^{-13}$ & $-6.174(49) \times 10^{-13}$ \\
	Re($A_2$) $64^3$ & $1.3410(89)\times 10^{-8}$& $1.4386(95)\times 10^{-8}$ \\
	Im($A_2$) $64^3$ & $-6.037(71)\times 10^{-13}$ & $-6.548(78)\times 10^{-13}$\\
	\hline
	\end{tabular}
		\caption{Comparison of matrix elements calculated with leading order (LO) and next-to-leading order (NLO) Wilson coefficients.}
	\label{tab:wcmel}
		\end{table}

To estimate the error in the Wilson coefficients, we compare the results for $A_2$ using Wilson coefficients calculated at leading order and next-to-leading order.
	We have used the set of coefficients evaluated in the \MSbar scheme at 3\,GeV, which are shown in Table~\ref{tab:wilson}~\cite{elaine}, and the standard parametrization of Wilson coefficients was used, i.e. $C_i=z_i+\tau y_i$ where $\tau$ is the ratio of CKM matrix coefficients $\tau = -\frac{V_{ts}^*V_{td}}{V_{us}^*V_{ud}}$. Throughout this paper we use the particle data group convention for the matrix elements, where $V_{us}=0.97425$, $V_{ud}=0.2252$ and $\tau=0.0014148-0.0005558 i$.
	The results for matrix elements calculated at leading and next-to-leading orders are shown in Table~\ref{tab:wcmel}.
	From the differences between the entries in the columns marked as LO and NLO we 
	estimate that the uncertainties are 6.8\% for Re($A_2$) on both sets of ensembles and 10\% (8\%)
for Im($A_2$) on the $48^3$ ($64^3$) ensembles.

Tables \ref{tab:errre} and \ref{tab:errim} show
our estimates of systematic errors associated with the
results for Re$(A_2)$ and Im$(A_2)$ presented in this
paper.  The evaluation of the continuum limit of $A_2$
is discussed in the following section.  As will be seen, the
systematic error associated with this extrapolation is
negligible with respect to the statistical errors.  
Consequently no discretization error is shown in
Tables \ref{tab:errre} and \ref{tab:errim}.
The values in the column marked ``Cont." are the errors assigned to our continuum-extrapolated
results, and are simply the larger of the corresponding entries from the $48^3$ and $64^3$ columns.
	We can see that the dominant contribution to the systematic error for both real and imaginary parts of $A_2$ on both ensembles comes from the uncertainty in Wilson coefficients.
	
	\begin{table}[t]
		\begin{tabular}{|l|c|c|c|}
			\hline
			Re$A_2$ systematic errors& $48^3$ & $64^3$ & cont.\\
			\hline
			\rm{NPR} (nonperturbative) & 0.1\% & 0.1\% & 0.1\% \\
			\rm{NPR} (perturbative) & 2.9\% & 2.5\% & 2.9\% \\
			Finite-volume corrections & 2.2\% & 2.4\% & 2.4\% \\
			Unphysical kinematics & 1.8\% & 4.5\% & 4.5\%\\
			Wilson coefficients & 6.8\% & 6.8\% & 6.8\% \\
			Derivative of the phase shift & 1.1\% & 0.6\% & 1.1\% \\
			\hline
			Total & 8\% &9\%  & 9\% \\
			\hline
		\end{tabular}
		\caption{Systematic error breakdown for Re\,$A_2$}
		\label{tab:errre}
	\end{table}
	\begin{table}[t]
		\begin{tabular}{|l|l|l|l|}
			\hline
			Im$A_2$ systematic errors& $48^3$ & $64^3$ & cont\\
			\hline
			\rm{NPR} (nonperturbative) & 0.1\% &0.1\% & 0.1\% \\
			\rm{NPR} (perturbative) & 7.0\% & 6.2\% & 7.0\% \\
			Finite-volume corrections & 2.4\% & 2.6\% & 2.6\% \\
			Unphysical kinematics & 0.2\% & 1.1\% & 1.1\% \\
			Wilson coefficients & 10\% & 8\% & 10\%\\
			Derivative of the phase shift & 1.1\%& 0.6\% & 1.1\% \\
			\hline
			Total & 12\% & 10\% & 12\% \\
			\hline
		\end{tabular}
		\caption{Systematic error breakdown for Im $A_2$}
		\label{tab:errim}
	\end{table}

\begin{table}[t] 
\centering
\begin{tabular}{c|c|c|c}
\hline\hline
& 48I & 64I & Phys.\\ 
\hline \hline
$m_{\pi}/m_{\Omega}$     &  0.08296(17) & 0.08220(19) & 0.08073\\
$m_{K}/m_{\Omega}$     &  0.29740(32) & 0.29982(37) & 0.29643\\
\end{tabular}
\caption{The ratios of the pion and kaon mass to the Omega baryon mass on the $48^3$ and $64^3$ ensembles as well as the physical value.\label{tab:massratios}}
\end{table}

\section{Continuum Extrapolation} \label{sec:contextrap}

In this section we discuss the extrapolation of the
results obtained on the $48^3$ and $64^3$ ensembles
to the continuum limit. We divide this discussion
into two parts. In the first we present the complete
physical results for the complex amplitude $A_2$ in
the continuum limit.  As we will observe, the dominant
error in our result comes from the perturbative error
assigned to the Wilson coefficients.  This may
be reduced in the future if higher order perturbation
theory results become available or if lattice step-scaling
methods are used to allow present perturbative results
to be applied at a higher energy scale.  Therefore, in
the second part we determine the continuum limit of the
individual matrix elements themselves, normalized in
the regularization-independent $(\s{q},\s{q})$ and
$(\gamma,\gamma)$ schemes.

\subsection{Continuum limit of Re$(A_2)$ and Im$(A_2)$}\label{subsec:contA2}
As already mentioned in Sec.\,\ref{sec:errors} the quark masses used in these ensembles are very slightly larger than their physical values.
This is illustrated in
Table~\ref{tab:massratios}, in which we compare the physical and
simulated values of the dimensionless quantities $m_\pi/m_\Omega$ and
$m_K/m_\Omega$, which are highly sensitive to the light- and heavy-quark
masses respectively.
In order to determine the values of the lattice spacing we must
therefore perform a short chiral extrapolation; this is achieved using a
simultaneous chiral and continuum ``global fit'' that incorporates data
from both ensembles. Since
the (renormalized) quark masses on the two ensembles are very similar,
we must include additional ensembles in order to have a sufficient
spread of masses for the determination of the chiral dependence. The full set of ensembles and details of this
procedure can be found in~\cite{Blum:2014tka}.

The determination of $A_2$ presented here was performed
using 76 configurations of the $48^3$ ensemble, whereas the lattice
spacings in~\cite{Blum:2014tka} were computed using 80. In order to
preserve the full correlations between the jackknife samples of $A_2$
and the corresponding superjackknife samples of the lattice spacing, we
repeated the global fit analysis using the same 76 configurations.
The details of the binning are also different. In~\cite{Blum:2014tka} we binned the $48^3$ data over 5 successive
measurements (100 MD time units) in order to take into account the
observed autocorrelations in the data, whereas in the present calculation, as explained in Sec.\,\ref{sec:simulations}, we construct 19 bins each of 4 configurations. 
These differences lead to determined values of the lattice spacings in Eq.\,(\ref{eq:a64a48}) below which are a little different from those in~\cite{Blum:2014tka}.
For the $64^3$ ensembles we use the same set of 40 configurations for the evaluation of $A_2$ and the same binning as in the global fit in~\cite{Blum:2014tka}.

In order to estimate the systematic errors  due to the chiral extrapolation and finite volume
in the determination of the lattice spacings, we have performed our fits using
three different chiral Ans\"{a}tze: NLO SU(2) chiral perturbation
theory, with and without finite-volume corrections (referred to as the
ChPTFV and ChPT forms respectively), and a linear Ansatz (referred to as
the ``analytic'' form). In practice we found the lattice spacings obtained
from all three Ans\"{a}tze to be consistent to within a fraction of the
statistical error due to the dominance of the near-physical data, hence
we treat these systematic errors as negligible. The final results for the values of the lattice spacing are
\begin{equation}\label{eq:a64a48}
a^{-1}_{\rm 64} = 2.3584(70)\ {\rm GeV}\quad\mathrm{and}\quad
a^{-1}_{\rm 48} = 1.7280(41)\ {\rm GeV}\,,
\end{equation}
where the errors are statistical only.

\begin{figure}[t]
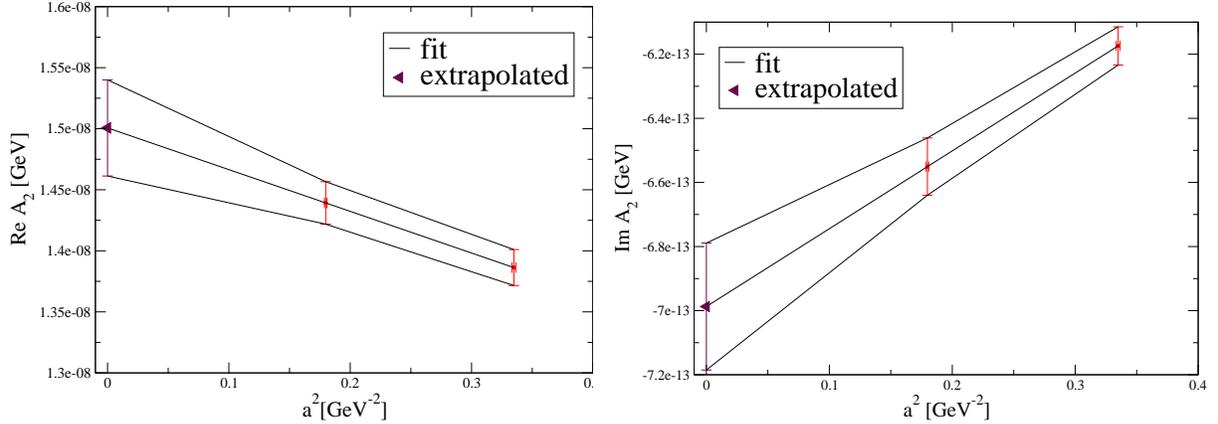

\centering
\includegraphics[width=0.48\textwidth]{fig9_extrap_ChPTFV_reim0_plot.eps}
\includegraphics[width=0.48\textwidth]{fig9_extrap_ChPTFV_reim1_plot.eps}
\caption{The continuum extrapolation of ${\rm Re}(A_2)$ (left) and ${\rm Im}(A_2)$ (right). The points at finite lattice spacing are taken from Tab.\,\ref{tab:a2int} for the $(\s{q},\s{q})$ intermediate renormalization scheme. \label{fig-A2contextrap} }
\end{figure}

The lattice matrix elements $M_i$ scale as $a^3$ and so small differences in the lattice spacing become amplified.
We have performed the continuum extrapolation of $A_2$ using the lattice spacings obtained with each of the three chiral Ans\"{a}tze; the extrapolated values are given in Table~\ref{tab-A2contvals}. In Fig.~\ref{fig-A2contextrap} we show the continuum extrapolation in the $(\slashed{q},\slashed{q})$ scheme using the lattice spacings obtained with the ChPTFV chiral Ansatz. We use results obtained with this Ansatz as our central values for each lattice spacing and for the extrapolated value in the continuum.

We obtain an estimate of the component of the chiral extrapolation error arising from the lattice spacing determination by taking the difference between the continuum values obtained using the ChPTFV and analytic lattice spacings. The full jackknife differences are $0.3(2.6)\times 10^{-10}$ and $0.1(1.2)\times 10^{-14}$ for the real and imaginary parts respectively. As with the lattice spacings, we cannot resolve these differences within the \rm{stat}istical error; hence we set the chiral error to zero. On the other hand the jackknife differences between the ChPTFV and ChPT Ans\"{a}tze are resolvable as they differ only in small Bessel function corrections and are thus highly correlated: we obtain $3.4(1.0)\times 10^{-11}$ and $1.59(47)\times 10^{-15}$ for the real and imaginary parts respectively. Nevertheless, these errors are only 5\%--8\%
of the statistical error and can therefore also be neglected. This leads to the result
\begin{equation}
{\rm Re}(A_2) = 1.501(39)\times 10^{-8}\ {\rm GeV}\,\qquad\mathrm{and}\qquad
{\rm Im}(A_2) = -6.99(20)\times 10^{-13}\ {\rm GeV}\,,\label{eq:A2context}
\end{equation}
where the errors are statistical. 

\begin{table}[t]
\centering
\begin{tabular}{l|cc}
\hline\hline
Ansatz     & ${\rm Re}(A_2)\ (\times 10^{-8}\ {\rm GeV})$ & ${\rm Im}(A_2)\ (\times 10^{-13}\ {\rm GeV})$ \\
\hline
ChPTFV     &  1.501(39)                                       &   -6.99(20)  \\
ChPT       &  1.494(38)                                       &   -6.96(19)   \\
analytic   &  1.494(43)                                       &   -6.96(21)   \\
\end{tabular}
\caption{The continuum values of ${\rm Re}(A_2)$ and ${\rm Im}(A_2)$ determined using the lattice spacings obtained with each of the three chiral Ans\"{a}tze.\label{tab-A2contvals} }
\end{table}

Our final result for $A_2$ is obtained by assigning
the 9\% and 12\% systematic errors from Tables~\ref{tab:errre} and \ref{tab:errim} 
as the systematic errors to be associated with the
values for Re$(A_2)$ and Im$(A_2)$ given in Eq.\,(\ref{eq:A2context}):
\begin{equation}
\boxed{{\rm Re}(A_2) = 1.50(4)_\mathrm{stat}(14)_\mathrm{syst}\times 10^{-8}\ {\rm GeV};\qquad
{\rm Im}(A_2) = -6.99(20)_\mathrm{stat}(84)_\mathrm{syst}\times 10^{-13}\ {\rm GeV}\,.}\label{eq:A2final}
\end{equation}

\begin{table}[t]
		\begin{center}
                \begin{tabular}{|c|c|c|c|}
                        \hline
                        Systematic errors in Im$A_2$/Re$A_2$& $48^3$ & $64^3$ & cont\\
                        \hline
                        \rm{NPR} (nonperturbative) & 0.1\% &0.1\% & 0.1\% \\
                        \rm{NPR} (perturbative) & 7.6 \% & 6.7 \% & 7.6 \% \\
                        Finite-volume corrections & 3.5 \% & 3.5 \% & 3.5 \% \\
                        Unphysical kinematics & 1.8 \% & 4.6\% & 4.6\% \\
                        Wilson coefficients & 12.0 \% & 10.5 \% & 12.0\%\\
                        Derivative of the phase shift &  0 & 0 & 0 \\
                        \hline
                        Total & 14.7\% & 13.7\% & 15.3\% \\
                        \hline
                \end{tabular}
                \caption{Systematic error breakdown for Im$A_2$/Re$A_2$.}
                \label{tab:errimre}
		\end{center}
        \end{table}

In order to estimate the unknown quantity Im$A_0$, we combine our results for $A_2$ with the experimental values of Re$A_0=3.3201(18)\times 10^{-7}\,$GeV and $\epsilon^\prime/\epsilon=(1.65\pm 0.26)\times 10^{-3}$~\cite{Agashe:2014kda}. To this end we start by evaluating the ratio Im$A_2$/Re$A_2$, taking into account any statistical correlations between the real and imaginary parts by performing the analysis within the jackknife procedure. On the two ensembles we find 
\begin{equation}
 \left(\frac{\mathrm{Im}A_2}{\mathrm{Re}A_2}\right)_{\!\!48^3}=-4.45(5)_{\mathrm{stat}}(65)_{\mathrm{syst}} \times 10^{-5}\quad\mathrm{and}\quad
\left(\frac{\mathrm{Im}A_2}{\mathrm{Re}A_2}\right)_{\!\!64^3}=-4.55(5)_{\mathrm{stat}}(62)_{\mathrm{syst}} \times 10^{-5}.
\end{equation}
The systematic errors for this ratio are given in Table \ref{tab:errimre}; they are generally combined in quadrature
except for that due to the derivative of the phase shift because the Lellouch-L\"uscher factor cancels in the ratio.
It is interesting to note that if instead of adding the errors in the Wilson coefficients for Re$A_2$ and Im$A_2$ in quadrature as in Table~\ref{tab:errimre}, we had calculated the ratios with the coefficients at leading and next-to-leading order respectively and taken the difference as a measure of the uncertainty we would have obtained a much smaller answer (3.6\% instead of 12\%). Since the operators which give the dominant contributions to the real and imaginary parts are different, and in the absence of an understanding which might suggest a correlation between their Wilson coefficients, we prefer to be cautious and take the larger uncertainty. We find a similar feature in the NPR perturbative error. 

The continuum extrapolation of the dimensionless ratio Re$A_2/\mathrm{Im}A_2$ is milder than that of Re$A_2$ and Im$A_2$ separately and we obtain
\begin{equation}\label{eq:ratiocontinuum}
\left(\frac{\mathrm{Im}A_2}{\mathrm{Re}A_2}\right)_{\!\!\mathrm{continuum}}=-4.67(72) \times 10^{-5}.
\end{equation}

Using this ratio, we can calculate the electroweak penguin contribution to $\epsilon'/\epsilon$, given by
\begin{equation}\label{eq:epeEWP}
	\left(\frac{\epsilon '}{\epsilon}\right)_{\mathrm{EWP}} \equiv \frac{\omega}{\sqrt 2 \left| \epsilon \right|}\frac{\mathrm{Im}A_2}{\mathrm{Re}A_2} = -6.6(10) \times 10^{-4},
\end{equation}
where we have used the values $\omega \equiv \frac{\mathrm{Re} A_2}{\mathrm{Re}A_0}=0.04454(12)$ and $\left| \epsilon \right| = 2.228(11)\times 10^{-3}$ from \cite{Blum:2012uk}. This value for $(\epsilon'/\epsilon)_{\mathrm{EWP}}$ is consistent with our previously quoted value $-6.25(44)(119)\times 10^{-4}$~\cite{Blum:2012uk}. Finally, for 
Im$A_0$ we find
\begin{equation}\label{eq:ImA0}
\mathrm{Im}A_0 = \mathrm{ReA_0}\,\left(\frac{\mathrm{Im}A_2}{\mathrm{Re}A_2}-\frac{\sqrt{2}|\epsilon|}{\omega}
\frac{\epsilon^\prime}{\epsilon}
\right)
= -5.40(64) \times 10^{-11}\,\mathrm{GeV}\,.
\end{equation}

	\begin{figure}[t]
		\begin{center}
		\includegraphics[scale=0.5]{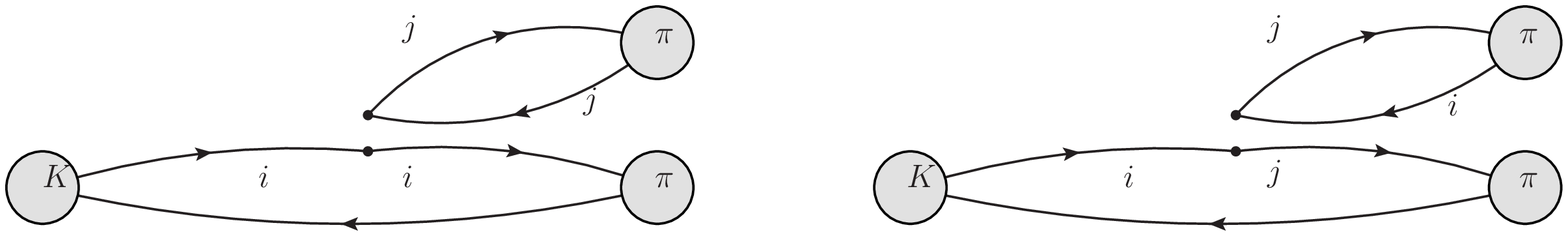}
		\caption{Dominant contractions contributing to Re($A_2$): $C_1$ (left) and $C_2$ (right).}
		\label{fig:kppi2}
		\end{center}
	\end{figure}

	\begin{figure}[t]
		\begin{center}
		\includegraphics[width=0.45\hsize]{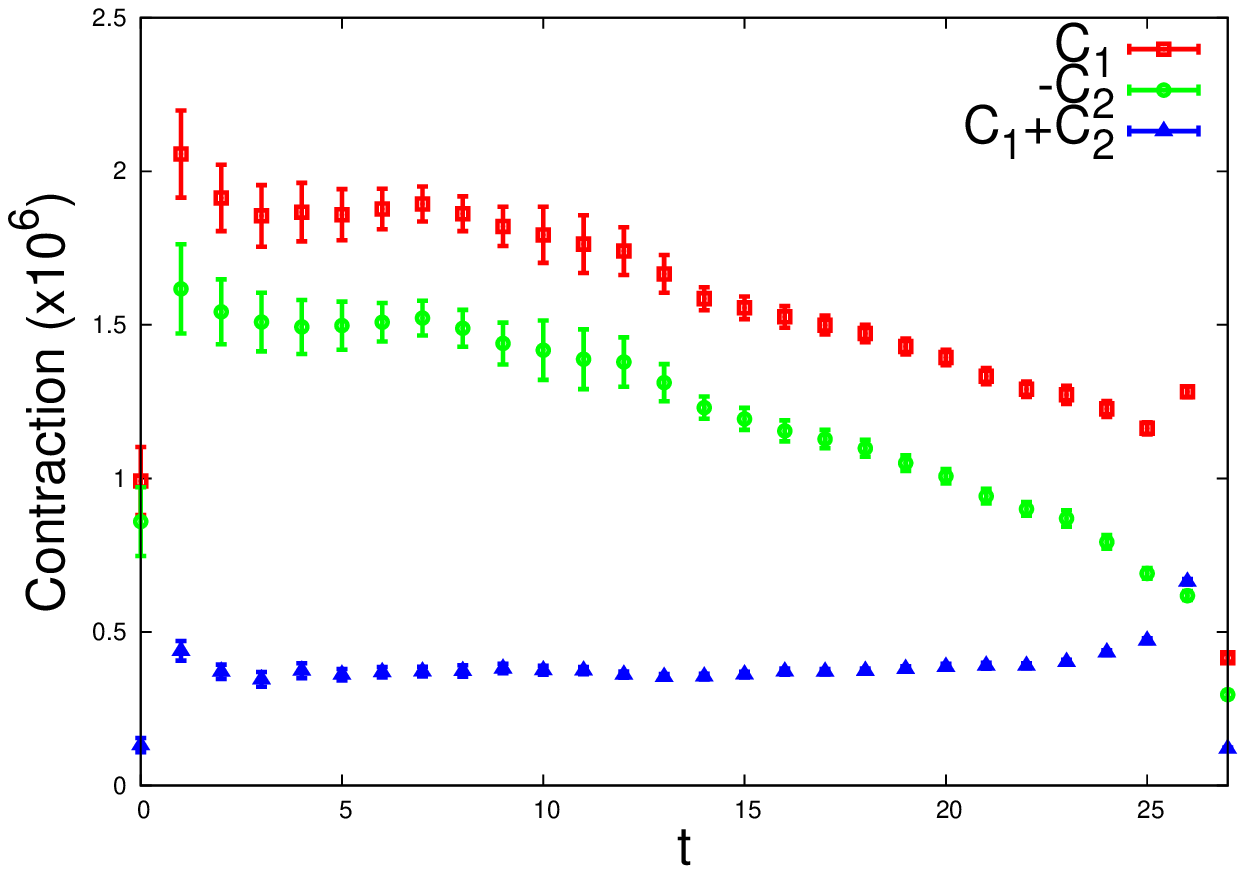}\qquad\quad\includegraphics[width=0.45\hsize]{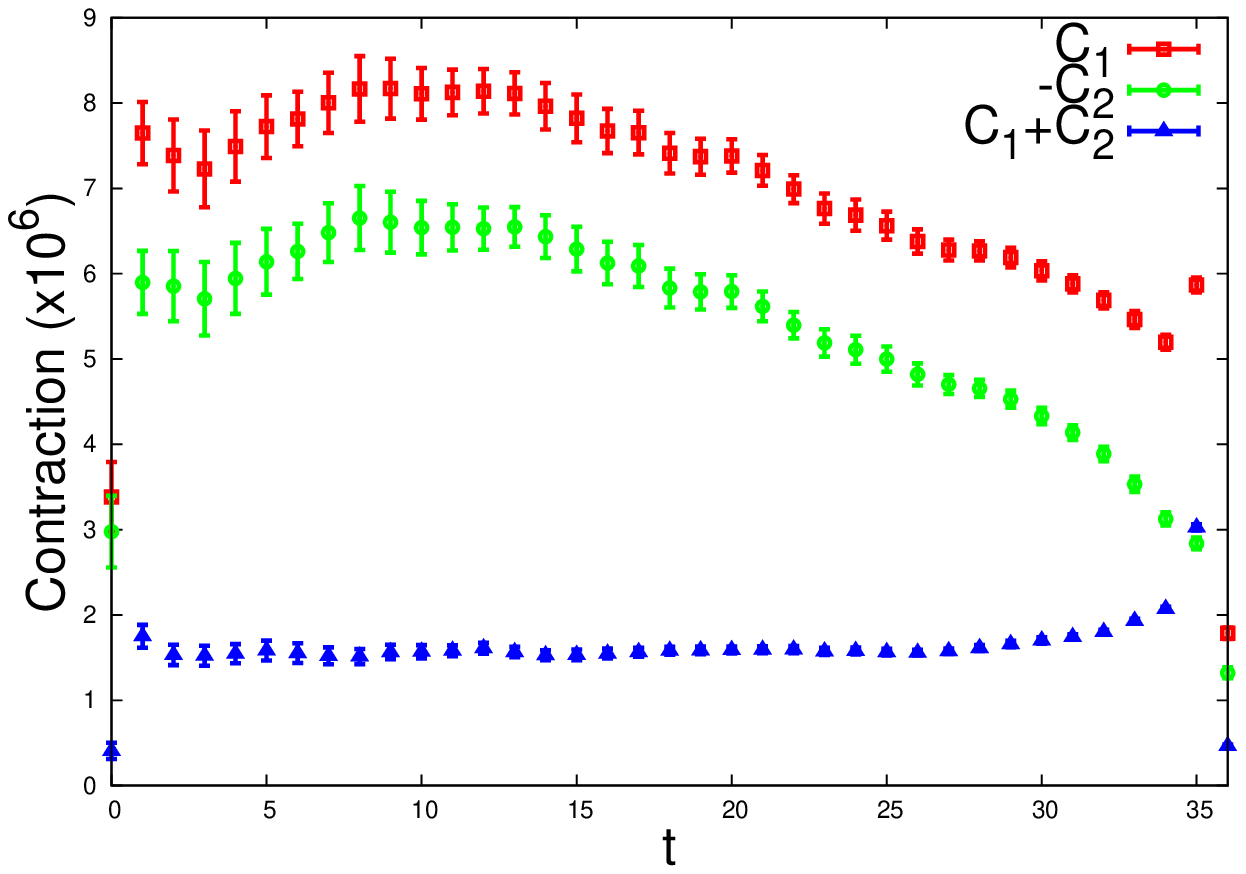}
		\caption{Cancellation of dominant contributions to Re($A_2$) on the $48^3$ ensembles with a $K$\,-\,$\pi\pi$ separation of 27 and the $64^3$ ensembles with separation 36.}
		\label{fig:c1c2}
		\end{center}
	\end{figure}
	
The results in Eqs.\,(\ref{eq:epeEWP}) and (\ref{eq:ImA0}) were obtained using our result for Im$A_2$/Re$A_2$ in Eq.\,(\ref{eq:ratiocontinuum}). If instead we take Im$A_2$ from our calculation, Eq.\,(\ref{eq:A2final}), and combine it with the experimental result Re$A_2=1.4787(31)\times 10^{-8}$\,GeV we obtain, Im$A_2$/Re$A_2=-4.73(58)\times 10^{-5}$, $(\epsilon^\prime/\epsilon)_\mathrm{EWP}=-6.69(82)\times 10^{-4}$ and Im\,$A_0=-5.42(63)\times 10^{-11}$\,GeV.
\emph{}
\subsection{Continuum limit of the RI-SMOM matrix elements}\label{subsec:merismom}
From the error budget in Table~\ref{tab:errimre} we see that the dominant uncertainty is due to the Wilson coefficients, which we take to be the difference between the leading and next-to-leading order contributions as defined in~\cite{Buchalla:1995vs}, where the calculations were based on \cite{Buras:1993dy,Ciuchini:1992tj,Ciuchini:1995cd}. In case the Wilson coefficients in the RI-SMOM schemes become known with better precision in the future, we present in Table~\ref{tab:mi-rismom} the $K^+\to\pi^+\pi^0$ matrix elements $M_i^{K^+}$ defined in Eq.\,(\ref{eq:mikdef}), with the operators $Q_i$ in Eqs.\,(\ref{eq:Q271})\,--\,(\ref{eq:Q88mx}) renormalized in the ($\s{q},\s{q}$) and ($\gamma,\gamma$) renormalization schemes at a renormalization scale of 3\,GeV. These matrix elements together with the new Wilson coefficients would enable an improved evaluation of $A_2$, without the need to recompute the matrix elements. 
The systematic errors for the (27,1) operator are estimated using the entries in Table~\ref{tab:errre} with the NPR(perturbative) and Wilson coefficient errors set to zero. This gives the errors of $2.8\%$, $5.1\%$ and $5.2\%$ for the $48^3$ and $64^3$ ensembles and in the continuum limit respectively. For the (8,8) operators using the entries in Table~\ref{tab:errim}, the same procedure leads to systematic errors of $2.6\%$, $2.9\%$ and $3.0\%$ for the $48^3$ and $64^3$ ensembles and in the continuum respectively.

For completeness we also convert these three $K^+\rightarrow (\pi\pi)_{I=2}$ matrix elements into those 
in the original 10 operator basis as defined in~\cite{Blum:2001xb}:
\begin{eqnarray}
M^{K^+}_{(27,1)}&=&3M^{K^+}_1 =3 M^{K^+}_2 = 2M^{K^+}_9=2M^{K^+}_{10}\\ 
M_{(8,8)}^{K^+}&=&2M^{K^+}_7\quad\textrm{and}\quad M_{(8,8)\textrm{mx}}^{K^+}=2M^{K^+}_8
\end{eqnarray}
where $M^{K^+}_i \equiv \langle (\pi\pi)_{I=2} \mid Q_i \mid K^+ \rangle$.

\begin{table}[t]
        \begin{center}
        \begin{tabular}{|c|c|c|c|c|}
        \hline
        Ensemble& Scheme&$M_{(27,1)}^{K^+}~ (\textrm{GeV}^3) $ & $M_{(8,8)}^{K^+}~(\textrm{GeV}^3)$ & $M_{(8,8)\textrm{mx}}^{K^+}~(\textrm{GeV}^3)$\\
        \hline
        $48^3$ &($\s{q},\s{q}$)  & $0.04761(39)(133) $ &$0.7026(52)(183) $ &$3.892(28)(101)$ \\
        $64^3$ & ($\s{q},\s{q}$)   & $0.04848(32)(247) $& $0.8412(88)(244) $& $4.140(44)(120)$\\	
        $48^3$ &($\gamma,\gamma$)  & $0.04473(37)(128) $ &$0.7112(53)(185)$ &$3.471(26)(90)$ \\
        $64^3$ & ($\gamma,\gamma$) & $0.04664(31)(238) $&0.8477(88)(246) &3.724(40)(108)\\ 
	\hline
	Continuum & ($\s{q},\s{q}$) & 0.0506(13)(26) & 1.003(22)(30) & 4.43(12)(13)\\
	Continuum & ($\gamma,\gamma$) & 0.0489(13)(25) & 1.007(23)(30) & 4.02(10)(12)\\
        \hline
        \end{tabular}
        \end{center}
        \caption{Results for the $K^+ \rightarrow (\pi\pi)_{I=2}$ matrix elements $M_i^{K^+}$ (defined in Eq.\,(\ref{eq:mikdef})) in two non-exceptional RI-SMOM renormalization schemes at the scale 3\,GeV. The first error is statistical, while the second one is the systematic uncertainty estimated as described in the text.}
        \label{tab:mi-rismom}
        \end{table}
\section{Conclusions}\label{sec:concs}
Before briefly summarizing our results and discussing prospects for future calculations we confirm our finding, first presented in~\cite{Boyle:2012ys}, that there is a significant cancellation between the two dominant contributions to Re\,$A_2$.  As explained above, Re($A_2$) is dominated by the matrix element of the $(27,1)$ operator and is proportional to the sum of the two contractions $C_1$ and $C_2$ in Fig.\,\ref{fig:kppi2}. While na\"ive factorization, frequently used for phenomenological estimates, suggests that $C_1=3\,C_2$ because of the color suppression in $C_2$, we find a strong cancellation between these two contributions. For the $48^3$ and $64^3$ ensembles studied in this paper, we illustrate this cancellation in Fig.\,\ref{fig:c1c2}. (In Sec.\,\ref{sec:bare} we explain that the numerical results in this paper were obtained from correlation functions with even values of $t_{\pi\pi}$. The choice of $t_{\pi\pi}=27$ for the 48~ensembles in Fig.\,\ref{fig:c1c2} is made to ensure that the cancellation is illustrated at the same value of $t_{\pi\pi}$ in physical units on the two sets of ensembles.)
As explained in ~\cite{Boyle:2012ys} we believe that this cancellation is a significant component in explaining the $\Delta I=1/2$~rule. Although we have not completed the calculation of $A_0$ at this stage, we note that the contributions of the $(27,1)$ operator all contribute with the same sign. 
A similar partial cancellation occurs between the two corresponding contractions in the evaluation of the $B_K$ parameter of neutral kaon mixing as pointed out in~\cite{Lellouch:2011qw} and subsequently confirmed in~\cite{Boyle:2012ys,Carrasco:2013jda}.

Our \textit{ab initio} determination of $A_2$ shows clearly that phenomenological approaches based on the dominance of na\"ive factorization are not consistent. We note however, that there were nonlattice studies based on chiral perturbation theory and the $1/N$ expansion, where $N$ is the number of colors, which indicated
that $C_2$ may have the opposite sign to $C_1$~\cite{Bardeen:1986vz,Pich:1995qp}. Of course, as illustrated in our results above, the $1/N$ expansion {\it per se} is not a good approximation; $C_2$ is suppressed by $1/N$ and yet is comparable to $C_1$. In different ways, the authors of~\cite{Bardeen:1986vz,Pich:1995qp} combine the expansion with leading short- and long-distance logarithms. In~\cite{Bardeen:1986vz} the authors use an Ansatz for matching the perturbative short-distance contributions and long-distance effects based on a chiral Lagrangian for mesons. In~\cite{Pich:1995qp} the authors compare the experimental value of Re$A_2$ with the leading term of the expansion to deduce that $C_2$ should be negative. For recent discussions of these two early approaches,
stimulated by our lattice QCD result [1, 2] and written
by subsets of their original authors, we refer the
reader to~\cite{Buras:2014maa,Pich:2014zta}.

Our earlier calculation of $A_2$ was performed on an ensemble at a single coarse lattice spacing, $a^{-1}=1.364$\,GeV~\cite{Blum:2011ng,Blum:2012uk}, and so not surprisingly the dominant systematic uncertainty was due to discretization errors. We estimated these to be about 15\%, although with only a single lattice spacing this could only be an estimate. In the present paper we repeat and refine the earlier calculation which is now performed on two finer ensembles with different lattice spacings, allowing for a continuum extrapolation.
We have determined Re\,$A_2$ to be $1.50(4)_{\rm{stat}}(14)_{\mathrm{sys}}\times 10^{-8}$\,GeV. This is 
consistent with the experimental values of $1.4787(31)\times 10^{-8}$ GeV from charged kaon decays
and $1.570(53)\times 10^{-8}$ GeV from neutral kaon decays.
We have also calculated the imaginary part of $A_2$ to be $-6.93(20)_{\rm{stat}}(84)_{\mathrm{sys}}\times 10^{-13}$~GeV, which was unknown until~\cite{Blum:2011ng,Blum:2012uk}. [We recall that the corresponding results from our earlier work were Re\,$A_2= 
1.38(5)_{\rm{stat}}(26)_{\mathrm{sys}}\times 10^{-8}$\,GeV and Im\,$A_2= 
-6.54(46)_{\rm{stat}}(120)_{\mathrm{sys}}\times 10^{-8}$\,GeV\,.]
Our results for Im and Re $A_2$ imply $(\epsilon^{\,\prime}/\epsilon)_{EWP} = -
6.6(10) \times 10^{-4}$. This can be compared to the result obtained via finite energy sum rules~\cite{Cirigliano:2002jy}, Re$(\epsilon^\prime/\epsilon)_{EWP}=-(11.0\pm 3.6)\times 10^{-4}$ (see also results based on vacuum saturation~\cite{Cirigliano:2002jy,Buras:2003zz}). We also mention for completeness that the continuum value of the two-pion phase shift is $\delta = -0.203(43)$.'

The errors are currently dominated by systematic uncertainties, the largest of which is due to the uncertainty in the (perturbative) evaluation of the Wilson coefficients (see Tables~\ref{tab:errre} and \ref{tab:errim}). 
It is testimony to the huge progress in the precision of lattice calculations that this is the case. We have aimed to be conservative in estimating this error, taking the difference between the lowest order and the next-to-lowest order as the uncertainty. The natural way to decrease this error is to perform higher-order perturbative calculations in the standard model but it may also be possible to use step scaling to increase the renormalization scale in the intermediate schemes (such as the RI-SMOM schemes used in this study) and hence to increase the scale at which the matching to the  \MSbar
scheme is performed and at which the Wilson coefficients are calculated.
It will be interesting to explore this possibility.

In order to have a fully quantitative understanding of the $\Delta I=1/2$ rule, to determine $\epsilon^\prime/\epsilon$ and to compare the result to the experimental value $\epsilon^\prime/\epsilon=(1.65\pm 0.26)\times 10^{-3}$ we need to perform the evaluation of $A_0$ at physical kinematics. 
A key ingredient which makes the calculation of $A_2$ feasible is the use of the Wigner-Eckart theorem described in Sec.\,\ref{sec:bare}. Together with the choice of volume and the use of antiperiodic boundary conditions for the $d$-quark in all three spatial directions, it ensures that the energy of the two-pion ground state is equal to $m_K$. Unfortunately this approach cannot be directly applied to the calculation of $A_0$; in particular the breaking of isospin symmetry by the boundary conditions invalidates the calculation. For example, the $\pi^0$ remains at rest with the antiperiodic boundary conditions, whereas the charged pions have nonzero momentum.
More sophisticated boundary conditions mixing
quarks and antiquarks and an isospin rotation, the so-called \textit{G-parity} boundary conditions~\cite{Wiese:1991ku,Kim:2003xt,Kim:2009fe,Kelly:2012eh,Kelly:2014usa},
must therefore be used instead for both the valence and the sea quarks. The evaluation of $A_0$ with G-parity boundary conditions is well underway and exciting progress has recently been reported in~\cite{ckdaiqian} and we anticipate the first complete calculation of $A_0$, albeit on a single lattice spacing, within the next year.

\section*{Acknowledgments}
The generation of the $48^3\times 96$ and $64^3\times128$ M\"obius DWF+Iwasaki ensembles used to calculate $A_2$ was performed
using the IBM Blue Gene/Q (BG/Q) Mira machine at the Argonne Leadership Class Facility
(ALCF) provided under the Incite Program of the U.S. DOE, on the STFC funded DiRAC
BG/Q system in the Advanced Computing Facility at the University of Edinburgh, and on the
BG/Q machines at Brookhaven National Laboratory (BNL). The DiRAC
equipment was funded by BIS National E-infrastructure Capital Grant No. ST/K000411/1, STFC Capital
Grant No. ST/H008845/1, and STFC DiRAC Operations Grants No. ST/K005804/1 and No. ST/K005790/1.
DiRAC is part of the National E-Infrastructure. Most of the measurements were also performed
on the DiRAC and Mira machines, with the remainder performed using the Brookhaven and the RIKEN-BNL Research
Center BG/Q computers at BNL.
The software used includes the CPS QCD code (\url{http://qcdoc.phys.columbia.edu/cps.html}), supported in part by the U.S. DOE SciDAC program, and the BAGEL (\url{http://www2.ph.ed.ac.uk/~paboyle/bagel/Bagel.html}) assembler kernel generator for high-performance optimized kernels and fermion solvers~\cite{Boyle:2009vp}. The gauge fixing for the 48I ensemble was performed using the CUTH cluster at Columbia University using the ``GLU" (Gauge Link Utility) codebase (\url{https://github.com/RJhudspith/GLU}).

T.B. is supported in part by the U.S. Department of Energy
Grant No. DE-FG02-92ER41989; N.H.C., R.D.M., D.Z. and H.Y.
by U.S. DOE Grant No.DE-SC0011941,
N.G. by Leverhulme Research Grant No. RPG-2014-118 and by the European 
Union under Grant No. 238353 (ITN STRONGnet); C.J., C.L. and
A. S. by U.S. DOE Contract No. DE-AC02-98CH10886(BNL) and T.J. and C.T.S. by UK STFC Grants No. ST/G000557/1 and No. ST/L000296/1.
C.K is supported by a RIKEN foreign postdoctoral research (FPR) grant.

\begin{appendix}
\section{$\boldsymbol{\beta(m_K,m_\pi,m_\pi)}$ and the Lellouch-L\"uscher Factor}
\label{app1}

In Sec.\,\ref{sec:FV} we use chiral perturbation theory to estimate the finite-volume corrections in our calculation of $A_2$ and consider the differences between the finite-volume sums and infinite-volume integrals in $\ell(m^2)$ and $\beta(q,m_1,m_2)$ defined in Eqs.\,(\ref{eq:ldef}) and (\ref{eq:betadef}). In the case with $q=(m_K,\vec{0})$ and $m_1=m_2=m_\pi$, $\beta(m_K,m_\pi,m_\pi)$ in Minkowski space has an imaginary part which leads to finite-volume corrections in Euclidean space which decrease only as inverse powers of the volume and not exponentially. These power corrections  are the one-loop chiral perturbation theory (NLO ChPT) contributions to the Lellouch-L\"uscher factor $F$ in Eqs.\,(\ref{eq:LLrelation}) and (\ref{eq:Fsq}). This factor is included fully in our analysis and so we must not include it again from NLO ChPT. A detailed study of how the Lellouch-L\"uscher factor arises in one-loop ChPT was performed in~\cite{Lin:2002nq}, but we hope that it will be useful to summarize the main points here.

In Minkowski space, performing the $k_0$ integration in the center-of-mass frame we obtain
\begin{equation}\label{eq:betamkmpimpimink}
\beta(m_K,m_\pi,m_\pi)=\int\frac{d^3\vec k}{(2\pi)^3}\,\frac{1}{\omega(\vec k)\,[m_K^2-4\omega^2(\vec k)+i\varepsilon]}\,,
\end{equation}
where $\omega^2(\vec k)=\left| \vec k\right| ^2+m_\pi^2$.
\begin{figure}[t]
\begin{center}
\includegraphics[width=0.5\hsize]{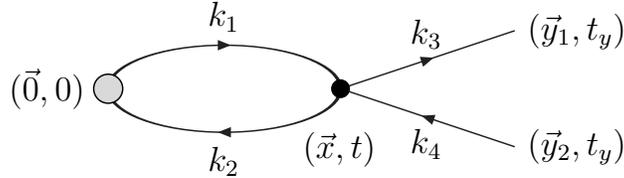}
\caption{Contribution to the correlation function in which two pions are produced by an operator at the origin (grey circle),
and rescatter by the strong interactions denoted by the filled circle. 
\label{fig:kpipi}}\end{center}
\end{figure}

In finite-volume Euclidean space we evaluate the correlation function illustrated in Fig.\,\ref{fig:kpipi}. The kaon propagator is irrelevant for our discussion and so we amputate it, and consider the two pions to be created at the origin, to rescatter and to be annihilated on the time slice at $t_y$. After performing the integrals over $\vec{y}_1$, $\vec{y}_2$ (with phase factors $e^{i\vec{q}\cdot\vec{y}_1}$ and $e^{-i\vec{q}\cdot\vec{y}_2}$ respectively) and $\vec{x}$ and exploiting the resulting $\delta$ functions, we obtain for this contribution to the correlation function:
\begin{equation}
I\equiv\int_{-\infty}^\infty dt\int\frac{d^3\vec k}{(2\pi)^3}\,\prod_{i=1}^4\frac{dE_i}{E_i^2+\omega_i^2}\,e^{i(E_1-E_2)t}\,
e^{i(E_3-E_4)(t_y-t)}\,,
\end{equation}
where in a finite volume the integral over $\vec{k}$ is replaced by the corresponding sum. Here $\omega_1^2=\omega_2^2= \omega^2(\vec k)=\left| \vec k \right|^2+m_\pi^2$ and $\omega_3^2=\omega_4^2=\omega^2(\vec q)=\left| \vec q\right|^2+m_\pi^2$ so that $\omega_{3,4}$ are not integration variables.

The energy integrals can now be performed by contour integration; there are three contributions depending on the value of $t$. 
\begin{enumerate}
\item The first contribution is from the interval $-\infty<t<0$ and gives
\begin{equation}\label{eq:term1}
I_1=\frac{e^{-2\omega(\vec q)\,t_y}}{32\,\omega^2(\vec q)}\int\frac{d^3\vec k}{(2\pi)^3}\,\frac{1}{\omega^2(\vec k)\,(\omega(\vec k)+\omega(\vec q))}\,.
\end{equation}
\item The second contribution comes from the region $0<t<t_y$ and gives
\begin{equation}\label{eq:term2}
I_2=\frac{e^{-2\omega(\vec q)\,t_y}}{32\,\omega^2(\vec q)}
\int\frac{d^3\vec k}{(2\pi)^3}\,\frac{1}{\omega^2(\vec k)\,(\omega(\vec k)-\omega(\vec q))}(1-e^{-2(\omega(\vec k)-\omega(\vec q))\,t_y})\,.
\end{equation}
\item Finally we have the contribution from the region $t_y<t<\infty$ which gives
\begin{equation}\label{eq:term3}
I_3=\frac{1}{32\,\omega^2(\vec q)}\int\frac{d^3\vec k}{(2\pi)^3}\,\frac{e^{-2\omega(\vec k)\,t_y}}{\omega^2(\vec k)\,(\omega(\vec k)+\omega(\vec q))}\,.
\end{equation}
\end{enumerate}

The contribution to the amplitude is given by the coefficient of \[\frac{e^{-2\omega(\vec q)t_y}}{4\,\omega^2(\vec q)}.\] In finite volume (FV) the integrals over $\vec k$ are replaced by the corresponding sums and we obtain the following three contributions. The first two are
\begin{equation}\label{eq:llterm1}
T_1 =\frac{1}{8L^3}\sum_{\vec{k}}\,\frac{1}{\omega^2(\vec k)\,(\omega(\vec k)+\omega(\vec q))}\,
\end{equation} 
from the region $t<0$, and
\begin{equation}\label{eq:llterm2}
T_2=\left(\frac{\nu_qt_y}{L^3}\right)\frac{1}{4\,\omega^2(\vec q)}+\frac{1}{8L^3}\sum_{|\vec k|\neq|\vec q|}\frac{1}{\omega^2(\vec k)
(\omega(k)-\omega(q))}\,,
\end{equation}
from the region $0<t<t_y$, 
where $\nu_q$ is the degeneracy of states with $\vec{k}=\vec{q}$. The term proportional to $t_y$ is the FV correction to the two-pion energy and it can be checked that this is correctly given by the L\"uscher quantization condition~\cite{Lin:2002nq}. Finally from the region $t_y<t<\infty$ we have
\begin{equation}
T_3=\left(\frac{\nu_q}{L^3}\right)\frac{1}{16\,\omega^3(q)}\,.
\end{equation}

We now separate the terms with $|\vec k|=|\vec q\,|$ from those where $|\vec k|\neq|\vec q\,|$. When $|\vec k|=|\vec q\,|$, we find a contribution
\begin{equation}
\frac{\nu_q}{L^3}\frac{1}{4\omega^2(\vec q)}\left\{\frac1{4\omega(\vec q)}+\frac1{4\omega(\vec q)}\right\}\,,
\end{equation}
where the first term in the braces corresponds to $T_1$ and the second corresponds to $T_3$. The contribution from $T_3$ is cancelled by the FV correction to the matrix element of the two-pion interpolating operator at $t_y$~\cite{Lin:2002nq} whereas the one from $T_1$ is a contribution to the FV effects in the amplitude.

The contributions from $|\vec k|\neq|\vec q\,|$ come from $T_1$ and $T_2$ and can be combined to give
\begin{equation}\label{eq:s1p}
\frac{1}{4\,L^3}\sum_{|\vec k|\neq|\vec q|}\frac{1}{\omega(\vec k)\,(\omega^2(\vec k)-\omega^2(\vec q))}\,.
\end{equation}
Thus in Euclidean finite volume we obtain 
\begin{equation}
S_1^\prime+\frac{\nu_q}{16L^3E^3}\,,
\end{equation}
where it is convenient to define
\begin{equation}
S_n^\prime=\frac{\omega^{n-1}(\vec q)}{4L^3}\sum_{|\vec k|\neq|\vec q|}\frac{1}{\omega^n(\vec k)\,(\omega^2(\vec k)-\omega^2(\vec q))}
\end{equation}
and the corresponding integrals by
\begin{equation}
J_n=\frac{\omega^{n-1}(\vec q)}{4}{\cal P}\int\frac{d^3 \vec k}{(2\pi)^3}\,\frac{1}{\omega^n(\vec k)\,(\omega^2(\vec k)-\omega^2(\vec q))}\,.
\end{equation}

Relating this sum to the corresponding integral gives the Lellouch-L\"uscher factor~\cite{Lin:2002nq}. We now make this more specific and determine the exponentially small corrections. In the difference $S_1^\prime-S_0^\prime$ there is no term with a pole at $\omega(k)=\omega(q)$ so that this difference can be related to the corresponding integral using the Poisson summation formula and the exponentially small finite-volume corrections can be identified:
\begin{eqnarray}
S_1^\prime-S_0^\prime&=&-\frac{1}{4L^3\omega(\vec q)}\sum_{|\vec k|\neq|\vec q|}\frac{1}{\omega(\vec k)(\omega(\vec k)+\omega(\vec q))}\\ 
&=&-\frac{1}{4L^3\omega(\vec q)}\sum_{\vec k}\frac{1}{\omega(\vec k)(\omega(\vec k)+\omega(\vec q))}+\frac{\nu_q}{8L^3\omega^3(\vec q)}\nonumber\\ 
&=&J_1-J_0+\frac{\nu_q}{8L^3\omega^3(\vec q)}+e_{1,0}\,.
\end{eqnarray}
Thus we see that the finite-volume and infinite-volume results are related by 
\begin{equation}\label{eq:aux1}
S_1^\prime+\frac{\nu_q}{16L^3E^3}=J_1-J_0+S_0^\prime+\frac{3\nu_q}{16L^3E^3}+e_{1,0}\,,
\end{equation}
where $e_{1,0}$ represent the exponentially small corrections,
\begin{equation}
e_{1,0}=-\frac{1}{8\pi^2\omega(\vec q)L}\sum_{\vec{n},n\neq 0}\frac{1}{n}\int_0^\infty k\,dk\,\frac{\sin(nkL)}{\omega(k)(\omega(k)+\omega(q))}\,,
\end{equation}
and $n$ and $k$ are $|\vec{n}\,|$ and $|\vec{k}\,|$ respectively. It was shown in \cite{Lin:2002nq} that $-J_0+S_0^\prime+\frac{3\nu_q}{16L^3E^3}$ is precisely the one-loop contribution to the Lellouch-L\"uscher factor. The residual exponentially small finite-volume effects are given by $e_{1,0}$. (The ultraviolet divergence cancels in the difference $J_0-S_0^\prime$, but if the zeta function regularization is used, as in \cite{Lellouch:2000pv}, then $J_0=0$.)
 
We have presented the above detailed discussion because we believe that there is a misunderstanding in the literature. In Eqs.\,(71) and (73) of~\cite{Aubin:2008vh}, the authors take the finite-volume corrections in $\beta(m_K,m_\pi,m_\pi)$ in Euclidean space to be the difference between the momentum integral and the corresponding sum over the integrand in Eq.\,(\ref{eq:betamkmpimpimink}) but with the replacement $m_K^2-4\omega^2(k)\to m_K^2+4\omega^2(k)$ in the denominator. Since there would now be no singularity in the denominator, the finite-volume corrections would be exponential and there would be no Lellouch-L\"uscher factor. The above derivation demonstrates instead the origin of the power corrections in the volume.
 
Throughout  the above discussion we assumed periodic boundary conditions in all three spatial directions so that $k_i=n_i\times (2\pi/L)$ where $n_i$ is an integer. In our determination of $A_2$ we use antiperiodic boundary conditions in all three directions so that 
\begin{equation}
e_{1,0}=-\frac{1}{8\pi^2\omega(q)L}\sum_{\vec{n},n\neq 0}\frac{(-1)^{n_x+n_y+n_z}}{n}\int_0^\infty k\,dk\,\frac{\sin(nkL)}{\omega(k)(\omega(k)+\omega(q))}\,.
\end{equation}
\end{appendix}

\end{document}